\RequirePackage{fix-cm}
\let\avec\vec
\documentclass[smallcondensed]{svjour3}     
\smartqed  
\usepackage{graphicx}
\graphicspath{{figures/}}
\usepackage{natbib}
\usepackage{mathtools}
\usepackage{amssymb,mathrsfs}
\usepackage{multirow}
\usepackage[linkcolor=blue,citecolor=blue,colorlinks]{hyperref}
\usepackage{txfonts}
\usepackage{stmaryrd}
\allowdisplaybreaks
\DeclareMathAlphabet{\mathpzc}{OT1}{pzc}{m}{it}

\newcommand{\e}{\mathrm{e}}
\newcommand{\ii}{\mathrm{i}}

\newcommand{\trans}[1]{{#1}^{\mathrm T}}

\newcommand{\hvec}[1]{\hat{\vec {#1}}}
\newcommand{\mat}[1]{\vec{#1}}
\newcommand{\sca}[1]{{\rm #1}}
\newcommand{\hmat}[1]{\hat{\mat {#1}}}

\newcommand{\dvec}[1]{\vec{\dot {#1}}{\vphantom{#1}}}

\newcommand{\op}[1]{\hat{\mathcal #1}}
\newcommand{\dd}{\mathrm{d}}

\renewcommand{\skew}{\mathrm{skew}}
\newcommand{\SO}{\mathrm{SO}_3}
\newcommand{\Id}{\mathbb{I}}

\DeclareMathOperator{\trace}{Tr}

\renewcommand{\equiv}{\coloneqq}
\defcitealias{Boue17}{Paper I}

\begin{document}

\title{Cassini states of a rigid body with a liquid core}

\author{Gwena\"el BOU\'E}

\authorrunning{G. Bou\'e} 

\institute{Gwena\"el Bou\'e  \at
              \email{gwenael.boue@observatoiredeparis.psl.eu}
           \and
           Sorbonne Universit\'e, PSL Research University, Observatoire de Paris, Institut de m\'ecanique c\'eleste et de calcul des \'eph\'em\'erides, IMCCE, F-75014, Paris, France
}

\date{Received: date / Accepted: date \\ \large Version: \today}

\graphicspath{ {figures/} }

\maketitle

\begin{abstract}
The purpose of this work is to determine the location and stability of the Cassini states of a
celestial body with an inviscid fluid core surrounded by a perfectly rigid mantle.
Both situations where the rotation speed is either non-resonant or trapped in a $p\!:\!1$ spin-orbit
resonance where $p$ is a half integer are addressed. The rotation dynamics is described by the
Poincar\'e-Hough model which assumes a simple motion of the core.  The problem is written in
a non-canonical Hamiltonian formalism.  The secular evolution is obtained without any truncation in
obliquity, eccentricity nor inclination. The condition for the body to be in a Cassini state is
written as a set of two equations whose unknowns are the mantle obliquity and the tilt angle of the
core spin-axis. Solving the system with Mercury's physical and orbital parameters leads to a maximum
of 16 different equilibrium configurations, half of them being spectrally stable. In most of these
solutions the core is highly tilted with respect to the mantle. The model is also applied to Io and
the Moon.

\end{abstract}


\section{Introduction}
\label{sec.intro}
The knowledge of the rotation motion of a rigid body allows to probe its internal structure, to
estimate some of its physical parameters and in some cases to constrain its past evolution. In that
scope, accurate observations are required as well as a dynamical model adapted to the problem
under study. Not all bodies are equal regarding to this technique as much information is gained
when the rotation is resonant. As an archetype, the Moon is characterised by a double synchronisation
described by Cassini's three famous laws \citep{Cassini93, Tisserand91, Beletsky01}. Using updated
values\footnote{\url{https://nssdc.gsfc.nasa.gov/planetary/factsheet/moonfact.html}} they read as 1/ the Moon
rotates uniformly around an axis that remains fixed in the body frame; the axial rotation period of
$27\,\mathrm{d}\ 7\,\mathrm{h}\ 43\,\mathrm{min}\ 14\,\mathrm{s}$ coincides with the period of
orbital revolution around the Earth. 2/ The Moon's equatorial plane has a constant inclination of
$6.68^\circ$ to the orbit plane, and of $1.54^\circ$ to the ecliptic plane. 3/ At all times the
Moon's spin axis lies in the plane formed by the normal to its orbit and the normal to the ecliptic
plane. 

Cassini's first law tells that the Moon is in a synchronous rotation state or, equivalently, in a
1:1 spin-orbit resonance. The remaining ones describe a secular spin-orbit resonance where the precession
frequency of the rotation axis equals the regression rate of the orbital ascending node. These two
commensurabilities are independent of each other. Indeed, \citet{Colombo66} proved the existence of the
second for axisymmetric bodies with arbitrary spin rate and, later, \citet{Peale69} generalised the
problem to triaxial bodies in any $p$\,:1 spin-orbit resonance where $p$ is a half integer.
In both studies, the phase space associated with the long term evolution of the rotation axis
possesses either two or four relative equilibria satisfying Cassini's last two laws. Accordingly,
they are nowadays referred to as Cassini states.

It should be noted that in real situations orbital planes do not precess uniformly. Their motion is
a combination of eigenmodes, each of them being associated with a given proper frequency. Therefore,
the spin axis cannot be in resonance with all these frequencies at the same time. Nevertheless, each
orbital harmonic does appear in the frequency decomposition of the spin axis motion. The body is
said to be in a Cassini state if its spectrum only contains orbital frequencies or linear
combinations of those with integer coefficients. In that case, the harmonic with the highest
amplitude (or obliquity) in the spin evolution can generally be identified with a relative equilibria
of the simplified problem where the orbital precession is reduced to that of a single eigenmode --
which is not necessarily the one with the highest amplitude (or inclination). The other orbital
frequencies generate small amplitude librations in the vicinity of the exact stable stationary points.

The solar system presents a variety of Cassini states. As said above, the Moon is in one of those
and has a synchronous rotation \citep{Cassini93}. This configuration is shared with most of the
regular satellites of the giant planets. Mercury also is in a Cassini state but its rotation is in a
3:2 resonance with its orbit \citep{Peale69}. Saturn is an example of asynchronous axisymmetric
planet whose spin axis is in secular spin-orbit resonance with the 2nd orbital harmonic ranked by
amplitude \citep{Ward04}. Finally, Jupiter is suspected to be in a Cassini state with its 3rd
orbital harmonic \citep{Ward06}. It can be stressed that due to the absence of dissipative process,
Saturn and Jupiter do not lie at the exact location of their respective Cassini states, a free
libration persists.

Precise measurements of the orientation of Cassini states led to several applications in the solar
system. In particular we can cite the determination of bounds on the dynamical ellipticities of the
Moon and of Mercury \citep{Peale69}; the inference of the presence of a liquid core inside Io
\citep{Henrard08} and of the existence of a global ocean beneath Titan's surface \citep{Bills11,
Baland11, Baland14, Noyelles14b, Boue17}; constraints on the past history of the outer solar system
\citep{Hamilton04, Boue09b, Vokroulicky15, Brasser15} and more specifically of Pluto satellite system
\citep{Quillen18}.

The goal of this work is to generalise the study made by \citet{Peale69} to the case of a rigid body with
a liquid core. This internal structure is a common model that has been used to interpret the rotation
of Mercury \citep{Peale76, Dufey09, Noyelles10, Peale14}, of the Moon \citep{Williams01, Meyer11} 
and of Io \citep{Henrard08, Noyelles12}. The presence of a fluid core has also been proposed to be
responsible for a resurfacing of Venus and for geophysical events in the past history of the Earth
\citep{Touma01}. Besides, \citet{Joachimiak12} considered the possibility that a liquid core within a
neutron star could mimic the signal of a planetary system.

The Cassini problem has been fully solved for entirely rigid bodies either under the gyroscopic
approximation \citep{Peale69} or without this hypothesis \citep{Bouquillon03}. In contrast, when
the body possesses a liquid core, the problem has only been studied in the vicinity of low
obliquities \citep[e.g.,][]{Touma01, Henrard08, Dufey09, Noyelles10, Noyelles12, Peale14}. Yet, recently
\citet{Stys18} obtained a set of equations truncated in eccentricity but valid at all obliquity
whose solutions provide the location of all the Cassini states of a three-layered body composed of a
rigid mantle, a fluid outer core and a rigid inner core. This analysis applied to the Moon has been
used to infer the orientation of its inner core ({\em ibid}).

In this paper, we analyse the Cassini states of a hollow rigid body filled with a perfect fluid
using the Poincare-Hough model \citep{Poincare10, Hough95} assuming a low ellipticity of the cavity.
The study is performed in a non-canonical Hamiltonian formalism exploiting as much as possible the
properties of rotations and of spin operators \citep{Boue06, Boue09, Boue16, Boue17b, Boue17, Boue19}. For
convenience, the Hamiltonian and the equations of motion are given in terms of matrices as in
\citep{Ragazzo15, Ragazzo17}, but the coordinate system in which these matrices are written is
arbitrary and only chosen at the end of the calculation. The equations defining the location of the
Cassini states match those derived in \citep{Stys18} when the moments of inertia of the rigid core
are artificially set to zero. We chose to restrain our study to 2-layered bodies because the full
phase space of this problem is still an unexplored territory. Moreover it allows to keep a model
with few degrees of freedom whose fixed points can easily be computed with Maple's package {\tt
RootFinding}.

The paper is organised as follows: Section~\ref{sec.notation} presents the model and the notation
used throughout the paper. The Hamiltonian of the problem and the equations of motion are 
obtained and averaged over the mean anomaly in Section~\ref{sec.hamiltonian}. The set of equations
defining the location of the Cassini states are given in Section~\ref{sec.Cassini}. Their
stabilities are studied in Section~\ref{sec.stability}. In Section~\ref{sec.results}, we apply the
model to Mercury, Io and the Moon. Finally we conclude in Section~\ref{sec.conclusion}.

\section{Model and notation}
\label{sec.notation}
\subsection{Notation}
In this work, we follow a notation close to that of \citet{Ragazzo15,Ragazzo17}, i.e., all matrices --
except the identity matrix $\Id$ -- are written in boldface, and to any vector $\avec x = (x_1, x_2, x_3)
\in \mathbb{R}^3$, we associate a skew-symmetric matrix $\mat x \in \skew(3)$ (represented by the
same letter) defined as
\begin{equation}
\mat x = \begin{bmatrix}
0 & -x_3 & x_2 \\
x_3 & 0 & -x_1 \\
-x_2 & x_1 & 0
\end{bmatrix}\;.
\end{equation}
The scalar product $\langle \cdot,\cdot\rangle$ between two matrices adopted in this work is slightly
different from that in \citep{Ragazzo15,Ragazzo17}\footnote{In \citet{Ragazzo15,Ragazzo17}, the
scalar product between two matrices is set to be $\langle \mat A, \mat B \rangle =
\trace(\trans{\mat A}\,\mat B)$ implying $|\mat x|^2=2|\avec x|^2.$}. We set $\langle \mat A, \mat B
\rangle = \frac{1}{2}\trace(\trans{\mat A}\,\mat B) = \frac{1}{2}\sum_{ij} A_{ij} B_{ij}$. This
choice implies that the norm of a skew-symmetric matrix is equal to that of its counterpart vector\,:
$|\mat x| = |\avec x\,|$. In particular, the skew-symmetric matrix of a unit vector is also unit. We denote
by $(\mat i, \mat j, \mat k)$ the canonical base frame matrices corresponding to the canonical base
frame vectors $(\avec i, \avec j, \avec k\,)$. Let us recall that the skew-symmetric matrix
associated with the vector product $\avec a\times \avec b$ is the commutator $[\mat a, \mat b] =
\mat a \mat b - \mat b \mat a$.


\subsection{Orbital motion}
We consider a rigid body with a
fluid core orbiting a central point mass $m_0$. We denote by $a$,
$e$, $i$, $v$, ${\cal M}$, $\varpi$, $\Omega$ the classical Keplerian elements, namely the semimajor
axis, the eccentricity, the inclination with respect to an invariant plane, the true anomaly, the
mean anomaly, the longitude of periapsis, and the longitude of the ascending node, respectively. The
system is supposed to be perturbed in such a way that the orbital plane is precessing uniformly with
constant inclination at the rate $\dot \Omega = g < 0$ around the normal $\avec k_L$ of the invariant
plane (also called Laplace plane). We define $n = \dot {\cal M}$ as the anomalistic mean motion.
Hereafter, the rigid body is assumed to be in a $p\!:\!1$ spin-orbit resonance, with $p$ being a
half integer. In that scope, we introduce a quantity associated with the rotation speed of the body,
namely
\begin{equation}
\omega_p \equiv pn + \dot\varpi - g \,.
\label{eq.omegap}
\end{equation}
The base frame vectors of the orbit are denoted ($\avec i_o, \avec j_o, \avec k_o$) with $\avec k_o$ along the
angular momentum and $\avec i_o$ at the intersection of the invariant plane with the orbit plane in the
direction of the ascending node. Let $\avec x(t)$ be the radius vector of norm $r(t)$ representing the position of the
body on its orbit. In the invariant frame $\kappa$, $\avec x(t)$ is given by
\begin{equation}
\avec x(t) = \mat R_3(\Omega) \mat R_1(i) \mat R_3(v + \varpi - \Omega) \begin{pmatrix} r(t) \\ 0 \\ 0 \end{pmatrix}\ ,
\end{equation}
where $\mat R_1$ and $\mat R_3$ are rotation matrices defined as
\begin{equation}
\mat R_1(\varphi) = \begin{bmatrix}
1 & 0 & 0 \\
0 & \cos\varphi & -\sin\varphi \\
0 & \sin\varphi &  \cos\varphi
\end{bmatrix} \ ,\qquad
\mat R_3(\varphi) = \begin{bmatrix}
\cos\varphi & -\sin\varphi & 0 \\
\sin\varphi &  \cos\varphi & 0 \\
0 & 0 & 1
\end{bmatrix}\ .
\end{equation}
From this radius vector, we define the symmetric traceless matrix $\mat S(t)$ by
\begin{equation}
\mat S(t) = \frac{a^3}{r(t)^5} \left(\avec x(t) \otimes \avec x(t) - \frac{r(t)^2}{3}\Id\right) \,.
\end{equation}

\subsection{Orientation}
The extended body is described by the Poincar\'e-Hough model \citep{Poincare10,Hough95}: it contains
a fluid {\em core} surrounded by a rigid layer now referred to as the {\em mantle}. The two
components are characterised by their inertia matrices $\mat I_{\rm c}$ and $\mat I_{\rm m}$,
respectively. Both matrices are assumed to be diagonal in the same frame $\mathrm K = (\,\avec I,
\avec J, \avec K\,)$, the latter being the principal frame of inertia of the whole body. We denote
by $\mat R\colon \mathrm K \rightarrow \kappa$ the rotation matrix defining the orientation of the
principal axes of inertia with respect to the invariant frame. We designate by $\mat \omega_{\rm m} =
\dot{\mat R}\trans{\mat R}$ the angular velocity matrix of the mantle relative to the invariant
frame and by $\mat \omega_{\rm c}$ the one associated with the {\em simple motion} of the liquid core (as
defined by \citet{Poincare10}) with respect to the invariant frame. Both angular velocities are written
in the invariant frame. A rotation matrix $\mat R_{\rm c}$ could be defined to record the simple motion of
the fluid core, but neither the kinetic energy nor the potential energy depend on this quantity.
Therefore, the state of the system is only given by the knowledge of $(\vec \omega_m, \vec \omega_c,
\mat R)$.

\subsection{Inertia matrices}
For each component $\lambda \in\{{\rm m},{\rm c}\}$, we consider the symmetric traceless matrix $\mat
B_\lambda$, such that the inertia matrix $\mat I_\lambda$ reads $\sca I_\lambda (\Id - \mat B_\lambda)$,
where $\sca I_\lambda = \frac{1}{3}\trace({\mat I_\lambda})$ is the mean moment of inertia of the
component $\lambda$. For the whole body, we equivalently define the overall inertia matrix $\mat I =
\sca I(\Id - \mat B)$, with $\sca I = \sca I_{\rm c} + \sca I_{\rm m}$ and $\mat B = (\sca I_{\rm c}\mat B_{\rm c} + \sca I_{\rm m}\mat B_{\rm m})/(\sca I_{\rm c}+\sca I_{\rm m})$.
$B$-matrices have been introduced by \citet{Ragazzo15, Ragazzo17}. For completeness we present
some of their properties. In an arbitrary frame, their expression in terms of Stokes coefficients
reads
\begin{equation}
\mat B_\lambda = \frac{2}{3}\frac{MR^2}{\sca I_\lambda} \begin{bmatrix}
3C^\lambda_{22} - \frac{1}{2}C^\lambda_{20} && 3S^\lambda_{\!22} && \frac{3}{2}C^\lambda_{21} \\[0.5em]
3S^\lambda_{\!22} && -3C^\lambda_{22}-\frac{1}{2}C^\lambda_{20} && \frac{3}{2}S^\lambda_{\!21} \\[0.5em]
\frac{3}{2}C^\lambda_{21} && \frac{3}{2}S^\lambda_{\!21} && C^\lambda_{20}
\end{bmatrix}\ ,
\end{equation}
where $M$ is the total mass of the body and $R$ its volumetric radius. This formulation suggests to
decompose the matrix $\mat B_\lambda$ into ``spherical harmonic matrices'' $\mat Y_{2m}$
\citep[see][]{Ragazzo15}. Here we define them (using the convention $S_{\!20} = 0$) such that
\begin{equation}
\mat B_\lambda = \frac{2}{3}\frac{MR^2}{\sca I_\lambda} \sum_{m=0}^2
\left(C^\lambda_{2m} \mat Y_{2m} + S^\lambda_{\!2m} \mat Y_{2-m} \right)\;,
\end{equation}
i.e.,
\begin{equation}
\begin{split}
&
\mat Y_{2-2} = \begin{bmatrix}
0 & 3 & 0 \\
3 & 0 & 0 \\
0 & 0 & 0
\end{bmatrix}
\ ,\quad
\mat Y_{2-1} = \begin{bmatrix} 
0 & 0 & 0 \\
0 & 0 & \frac{3}{2} \\
0 & \frac{3}{2} & 0
\end{bmatrix}
\ ,\quad
\mat Y_{20} = \begin{bmatrix} 
-\frac{1}{2} & 0 & 0 \\
0 & -\frac{1}{2} & 0 \\
0 & 0 & 1
\end{bmatrix}
\ , \\[0.5em]
&
\mat Y_{21} = \begin{bmatrix}
0 & 0 & \frac{3}{2} \\
0 & 0 & 0 \\
\frac{3}{2} & 0 & 0
\end{bmatrix}
\ ,\quad
\mat Y_{22} = \begin{bmatrix}
3 & 0 & 0 \\
0 & -3 & 0 \\
0 & 0 & 0
\end{bmatrix} \ .
\end{split}
\end{equation}
These spherical matrices are orthogonal. Indeed, a direct calculation shows that
\begin{equation}
\langle \mat Y_{2m}, \mat Y_{2k} \rangle = \frac{3}{4}\frac{1}{2-\delta_{m0}}\frac{(2+|m|)!}{(2-|m|)!}\delta_{mk} ,
\end{equation}
where $\delta_{mk} = 1$ if $m=k$ and $\delta_{mk} = 0$ otherwise. Let $\op R$ be the rotation operator associated with a rotation matrix $\mat R$ such that for any matrix $\mat A$, 
$
\op R [\mat A] \equiv \mat R \mat A \trans{\mat R} \,.
$
The rotation of the spherical harmonic matrices around the third axis takes a simple expression
\begin{equation}
\op R_3(\theta)[\mat Y_{2m}] 
 = \cos(m\theta) \mat Y_{2m} + \sin(m\theta) \mat Y_{2-m}\,.
\label{eq.R3Y2m}
\end{equation}
In the following, we need to compute the rotation around the first axis. The (less compact) result
reads
\begin{subequations}
\begin{eqnarray}
&&\op R_1(\theta)[Y_{2-2}] = 2\sin(\theta)\,\mat Y_{21} + \cos(\theta)\, \mat Y_{2-2} \ ,\\
&&\op R_1(\theta)[Y_{2-1}] = \frac{3}{2}\sin(2\theta)\,\mat Y_{20} + \frac{1}{4}\sin(2\theta)\,\mat Y_{22} + \cos(2\theta)\,\mat Y_{2-1}\ ,\\
&&\op R_1(\theta)[Y_{20}] = \frac{1+3\cos(2\theta)}{4}\mat Y_{20} -
\frac{1-\cos(2\theta)}{8}\mat Y_{22} - \frac{1}{2}\sin(2\theta)\,\mat Y_{2-1}\ ,\\
&&\op R_1(\theta)[Y_{21}] = \cos(\theta)\,\mat Y_{21} - \frac{1}{2}\sin(\theta)\,\mat
Y_{2-2}\ , \\
&&\op R_1(\theta)[Y_{22}] = -\frac{3}{2}(1-\cos(2\theta))\,\mat Y_{20} + \frac{3+\cos(2\theta)}{4}\mat Y_{22} - \sin(2\theta)\,\mat Y_{2-1}\ .
\end{eqnarray}%
\label{eq.R1Y2m}%
\end{subequations}%
Finally, we shall also introduce the multiplication table $[\mat Y_{2m}, \mat Y_{2k}]$ corresponding
to the commutator of any two spherical harmonic matrices. Given that each $\mat Y_{2m}$ is
symmetrical, these commutators are skew-symmetric, they can therefore be decomposed into the base
$(\mat i, \mat j, \mat k)$. The result is provided in Table~\ref{tab.mul}.

\begin{table}
\begin{center}
\caption{\label{tab.mul}Multiplication table of spherical harmonic matrices.}
\renewcommand{\arraystretch}{2.8}
\begin{tabular}{|c||c|c|c|c|c|} \hline
$[\cdot,\cdot]$ & $\mat Y_{2-2}$ & $\mat Y_{2-1}$ & $\mat Y_{20}$ & $\mat Y_{21}$ & $\mat Y_{22}$ \\ \hline\hline
$\mat Y_{2-2}$ & $\mat 0$ & $\displaystyle\frac{9}{2}\mat j$ & $\mat 0$ & $-\displaystyle\frac{9}{2}\mat i$ & $18\mat k$\\ \hline
$\mat Y_{2-1}$ & $-\displaystyle\frac{9}{2}\mat j$ & $\mat 0$ & $-\displaystyle\frac{9}{4}\mat i$ & $\displaystyle\frac{9}{4}\mat k$ & $-\displaystyle\frac{9}{2}\mat i$ \\ \hline
$\mat Y_{20}$ & $\mat 0$ & $\displaystyle\frac{9}{4}\mat i$ & $\mat 0$ & $-\displaystyle\frac{9}{4}\mat j$ & $\mat 0$ \\ \hline
$\mat Y_{21}$ & $\displaystyle\frac{9}{2}\mat i$ & $-\displaystyle\frac{9}{4}\mat k$ & $\displaystyle\frac{9}{4}\mat j$ & $\mat 0$ & $-\displaystyle\frac{9}{2}\mat j$ \\ \hline
$\mat Y_{22}$ & $-18\mat k$ & $\displaystyle\frac{9}{2}\mat i$ & $\mat 0$ & $\displaystyle\frac{9}{2}\mat j$ & $\mat 0$ \\ \hline
\end{tabular}
\end{center}
\end{table}

In our problem, inertia matrices $\mat I_\lambda$ are written in the principal frame of inertia,
therefore $B$-matrices only depend on $C_{20}$ and $C_{22}$. Furthermore, these matrices take an
even simpler form using $\alpha$ and $\beta$, respectively the polar and the equatorial flattening
coefficients as defined in \citet{Vanhoolst02} (see also Appendix~\ref{sec.flattening}),
\begin{equation}
\alpha_\lambda = -\frac{MR^2}{C_\lambda} C^\lambda_{20} \;,\qquad
\beta_\lambda  = 4\frac{MR^2}{C_\lambda} C^\lambda_{22} \;.
\end{equation}
We have, at first order in the flattening coefficients,
\begin{equation}
\mat B_\lambda = -\frac{2}{3}\alpha_\lambda \mat Y_{20} + \frac{1}{6}\beta_\lambda \mat Y_{22}\,.
\label{eq.BY2m}
\end{equation}

\section{Development of the Hamiltonian}
\label{sec.hamiltonian}
To model the evolution of the orientation of each component of the body driven by the orbital
motion, we start by writing the Lagrangian $L$ of the problem. The state of the
problem is described by the variables $(\mat \omega_{\rm m}, \mat \omega_{\rm c}, \mat R) \in
\skew(3)\times\skew(3)\times\SO$. At first order in the flattening coefficients, the expression of
the Lagrangian reads\footnote{For generic matrix $\mat A$ and vector $\avec x$, $$ \avec x \cdot
\mat A \avec x = \trace(\mat A)|\mat x|^2 - 2\langle \mat A, \mat x\trans{\mat x}\rangle\;.$$}
\begin{equation}
\begin{split}
L =&
  \frac{\sca I_{\rm m}}{2}\left(|\mat\omega_{\rm m}|^2+2\langle \mat R\mat B_{\rm m}\trans{\mat R}, \mat\omega_{\rm m}\trans{\mat\omega}_{\rm m}\rangle\right)
+ \frac{\sca I_{\rm c}}{2}\left(|\mat\omega_{\rm c}|^2+2\langle \mat R\mat B_{\rm c}\trans{\mat R}, \mat\omega_{\rm c}\trans{\mat\omega}_{\rm c}\rangle\right)
\\ &
+ 3\frac{\mathcal{G}m_0}{a^3}
 \sca I\langle\mat R\mat B\trans{\mat R}, \mat S(t)\rangle \;.
\end{split}
\end{equation}
In the first line, we recognise the kinetic energy of rotation of the mantle and of
the core. The last term provides the spin-orbit interaction.

The Poincar\'e-Lagrange equations for this system are \citep[e.g.,][]{Boue17}
\begin{subequations}
\begin{eqnarray}
&&
\frac{\dd}{\dd t} \frac{\partial {L}}{\partial \mat \omega_{\rm m}} = [ \mat \omega_{\rm m} ,
\frac{\partial L}{\partial \mat \omega_{\rm m}}] + \op J(L), \\
&&
\frac{\dd}{\dd t} \frac{\partial L}{\partial \mat \omega_{\rm c}} =
[ \mat \omega_{\rm c} , \frac{\partial L}{\partial \mat \omega_{\rm c}}] ,
\end{eqnarray}
\end{subequations}
where $\op J$ is the spin operator associated with the rotation $\mat R$ (see
Appendix~\ref{sec.spin} for a practical definition of the spin operator).
To proceed, we switch to the Hamiltonian formulation. Let $\mat \pi_\lambda$ be the angular momentum of
the mantle ($\lambda={\rm m}$) and of the core ($\lambda={\rm c}$). They are given by $\mat \pi_\lambda \equiv
\partial L/\partial \mat \omega_\lambda$, namely,
\begin{equation}
\mat \pi_\lambda =  \sca I_\lambda(\mat \omega_\lambda + \mat R\mat B_\lambda\trans{\mat R}\mat \omega_\lambda
                                     + \mat \omega_\lambda \mat R\mat B_\lambda\trans{\mat R})
\,,\qquad\lambda\in\{{\rm m},{\rm c}\}\,.
\end{equation}
We then perform a Legendre transformation to get the Hamiltonian $H$ of the problem:
$H =
\sum_\lambda \langle \mat \pi_\lambda , \mat \omega_\lambda \rangle - L$. Keeping only terms in
$\mathcal{O}(\mat B_i)$ in the kinetic energy, we get
\begin{equation}
\label{eq.Ham}
\begin{split}
H = &
 \frac{1}{2\sca I_{\rm m}}\left(|\mat\pi_{\rm m}|^2 - 2\langle\mat R\mat B_{\rm m}\trans{\mat R}, \mat \pi_{\rm m}\trans{\mat \pi}_{\rm m}\rangle\right)
+\frac{1}{2\sca I_{\rm c}}\left(|\mat\pi_{\rm c}|^2 - 2\langle\mat R\mat B_{\rm c}\trans{\mat R}, \mat \pi_{\rm c}\trans{\mat \pi}_{\rm c}\rangle\right)
\\ &
- 3\frac{\mathcal{G}m_0}{a^3}\sca I
 \langle\mat R\mat B\trans{\mat R}, \mat S(t)\rangle \;.
\end{split}
\end{equation}
The Poincar\'e-Hamilton equations read \citep{Boue17}
\begin{subequations}
\label{eq.motion}
\begin{eqnarray}
&&
\label{eq.pimdot}
\dot{\mat \pi}_{\rm m} =
 [\frac{\partial H}{\partial \mat \pi_{\rm m}}, \mat \pi_{\rm m}] - \op J (H) , 
\\ &&
\label{eq.pio}
\dot{\mat \pi}_{\rm c} = [\frac{\partial H}{\partial \mat \pi_{\rm c}}, \mat \pi_{\rm c}] ,
\\ &&
\dot{\mat R} = 
      \frac{\partial H}{\partial \mat \pi_{\rm m}} \mat R\,.
\label{eq.Cdot}
\end{eqnarray}
\end{subequations}
The equation of motion (\ref{eq.pio}) shows that the norm of $\vec\pi_{\rm c}$ is a constant of the
motion. In particular, if the fluid core is initially at rest ($\vec\pi_{\rm c} = \vec 0$), as long as there is no core-mantle friction (as in this model), the fluid remains at rest. For an alternative interpretation of this result, we refer the reader to \citep[and references therein]{Vanhoolst09}. Thus, the phase space of the problem is the manifold $\mathscr{M}$ of dimension 8 defined as
\begin{equation}
\mathscr{M} = \left\{(\mat \pi_{\rm m}, \mat \pi_{\rm c}, \mat R)\in 
\skew(3)\times\skew(3)\times\SO,  |\mat
\pi_{\rm c}| = c
\right\}
\end{equation}
where the constant $c\in\mathbb{R}_+$ constrains the norm of $\mat\pi_{\rm c}$.

\subsection{Change of variables}
Here, we are interested in the Cassini equilibrium configurations. These are the fixed
points of the spin axes in the frame rotating at the precession frequency of the orbital plane.
Moreover, we consider a body whose rotation speed is commensurate with the orbital mean motion (if
the system is not in spin-orbit resonance, one just needs to drop all resonant terms from the final
expressions).  In order to find the Cassini states, we perform a change of variables $(\mat \pi_{\rm
m},\mat \pi_{\rm c}, \mat R) \rightarrow (\mat \pi'_{\rm m}, \mat \pi'_{\rm c}, \mat R')$ closely related
to that made by \citet{Peale69}, namely,
\begin{subequations}
\begin{align}
&
\label{eq.pim'}
\mat \pi_{\rm m} = \mat R_3(\Omega)\mat R_1(i)\,\mat \pi'_{\rm m} \,\mat R_1(-i) \mat R_3(-\Omega)\,,
\\ &
\label{eq.pic'}
\mat \pi_{\rm c} = \mat R_3(\Omega)\mat R_1(i)\,\mat \pi'_{\rm c} \,\mat R_1(-i) \mat R_3(-\Omega)\,, 
\\ &
\label{eq.C'}
\mat R = \mat R_3(\Omega)\mat R_1(i)\,\mat R' \,\mat R_3(p{\cal M} + \varpi -\Omega) \;.
\end{align}%
\label{eq.piC'}%
\end{subequations}%
According to the transformation (\ref{eq.piC'}), the new momenta are expressed in the orbital frame
as is the rotation matrix $\mat R'$. But for the latter we also apply a rotation of angle
$(p{\cal M} + \varpi -\Omega)$ around the $K$-axis in order to precisely select the $p\!:\!1$
spin-orbit resonance. Let $H'$ be the Hamiltonian of the problem in the new variables.  Taking into
account the time-dependency of the change of variables (\ref{eq.piC'}), the equations of motion
(\ref{eq.motion}) remain unchanged\footnote{In the equations of motion written with the new set of
variables, the operator $\op J$ (representing derivatives with respect to $\mat R$) is replaced by
the equivalent operator $\op J'$ (expressing derivatives with respect to $\mat R'$).} if we choose
\begin{equation}
  H'(\pi'_{\rm m}, \mat\pi'_{\rm c}, \mat R') 
= H(\pi_{\rm m}, \mat \pi_{\rm c}, \mat R) 
- g \langle \mat k_L, \mat \pi'_{\rm m} + \mat \pi'_{\rm c}\rangle 
- \omega_p \langle \mat K', \mat \pi'_{\rm m} \rangle ,
\label{eq.Hprime}
\end{equation}
where $\omega_p$, given by Eq.~(\ref{eq.omegap}), is the angular velocity relative to the precessing
frame required to be at the exact $p\!:\!1$ spin-orbit resonance, and where $\mat K' = \mat
R'\mat k \mat R^{\prime\mathrm{T}}$ is the skew-symmetric matrix of the figure axis
of maximum inertia expressed in the orbit frame. We equivalently define $\mat I'$ and $\mat J'$ to
be equal to $\mat R' \mat i \mat R^{\prime\mathrm{T}}$ and $\mat R' \mat j \mat
R^{\prime\mathrm{T}}$, respectively. Notice that $\mat I'$ and $\mat J'$ are not anymore principal
axes of inertia because of the rotation $\mat R_3(p{\cal M}+\varpi-\Omega)$. To understand the
second term of the Hamiltonian (\ref{eq.Hprime}), let us recall that $g=\dot\Omega$ is the
precession rate of the orbital plane. Because the new variables are expressed in the orbit frame,
the normal of the invariant plane $\vec k_L$ must also be written in the orbit frame. Its
coordinates are
\begin{equation}
\avec k_L = \begin{pmatrix}
0 \\ \sin i \\ \cos i
\end{pmatrix} \quad\Rightarrow\quad
\mat k_L = \begin{bmatrix}
0 & -\cos i & \sin i \\
\cos i & 0 & 0 \\
-\sin i & 0 & 0 
\end{bmatrix}\;
.
\end{equation}
Naturally, the orbit base frame vectors $(\avec i_o, \avec j_o, \avec k_o)$ are now equal to $(\avec i,
\avec j, \avec k)$. Before writing the explicit expression of the new Hamiltonian, let us introduce
the following notation
\begin{subequations}
\begin{align}
&
\mat B'_\lambda(t) = \mat R_3(p{\cal M}+\varpi-\Omega) \mat B_\lambda \mat R_3(-p{\cal M}-\varpi+\Omega) , 
\qquad & \lambda\in\{{\rm m},{\rm c}\} \\
&
\mat S(t) = \mat R_3(\Omega)\mat R_1(i) \mat S'(t) \mat R_1(-i)\mat R_3(-\Omega) , &
\end{align}
\end{subequations}
and $\mat B' = (\sca I_{\rm m}\mat B'_{\rm m} + \sca I_{\rm c}\mat B'_{\rm c})/(\sca I_{\rm m}+\sca
I_{\rm c})$.  The different positions of the prime in the two equations are due to the fact that the
$B$-matrices are expressed in the frame fixed to the mantle while the $S$-matrix is written in the
invariant plane. Notice that the new $B$-matrices are now time-dependent (because of $\Omega$,
$\varpi$ and ${\cal M}$) as is the matrix $\mat S'$. For the seek of conciseness, in the sequel we
drop the explicit time-dependency in the notation. With these definitions, the new Hamiltonian reads
\begin{equation}
\begin{split}
\label{eq.Ham'}
H' = &
 \frac{1}{2\sca I_{\rm m}}\left(|\mat\pi'_{\rm m}|^2 - 2\langle\mat R'\mat B'_{\rm m}\trans{\mat R'}, \mat \pi'_{\rm m}\trans{\mat \pi'}_{\rm m}\rangle\right)
+\frac{1}{2\sca I_{\rm c}}\left(|\mat\pi'_{\rm c}|^2 - 2\langle\mat R'\mat B'_{\rm c}\trans{\mat R'}, \mat \pi'_{\rm c}\trans{\mat \pi'}_{\rm c}\rangle\right)
\\ &
- 3\frac{\mathcal{G}m_0}{a^3}\sca I
 \langle\mat R'\mat B'\trans{\mat R'}, \mat S'(t)\rangle
- g \langle\mat k_L,\,\mat \pi'_{\rm m} +\mat \pi'_{\rm c}\rangle
- \omega_p \langle \mat K', \mat \pi'_{\rm m} \rangle
\;.
\end{split}
\end{equation}

\subsection{Secular Hamiltonian}
At this stage, the two Hamiltonians (\ref{eq.Ham}) and (\ref{eq.Ham'}) are strictly equivalent. The
change of variables is valid whether the system is in a $p\!:\!1$ spin-orbit resonant or not. The
next step consists in averaging over the mean anomaly ${\cal M}$. In that scope, we make use of the
decomposition of the $B$-matrices in spherical harmonic matrices (Eq.~\ref{eq.BY2m}) and apply the
rule (\ref{eq.R3Y2m}) to perform a rotation of angle $\theta = (p{\cal M}+\varpi-\Omega)$ around the
third axis. We get
\begin{equation}
\mat B' = -\frac{2}{3}\alpha\mat Y_{20} + \frac{1}{6}\beta\,\Big(\cos(2p{\cal M}+2\varpi-2\Omega)\,\mat Y_{22}
                                                                +\sin(2p{\cal M}+2\varpi-2\Omega)\,\mat Y_{2-2}\Big) .
\label{eq.matB'}
\end{equation}
As for the $S$-matrix, we have
\begin{equation}
\mat S' = \left(\frac{a}{r}\right)^3 \mat R_3(v+\varpi-\Omega)\left(\frac{1}{6}\mat
Y_{22}-\frac{1}{3}\mat Y_{20}\right)\mat R_3(-v-\varpi+\Omega)\,,
\end{equation}
which can also be expanded using the formula (\ref{eq.R3Y2m}) into
\begin{equation}
\mat S' = \left(\frac{a}{r}\right)^3\left(-\frac{1}{3}\mat Y_{20}
+\frac{1}{6}\Big(\cos(2v+2\varpi-2\Omega)\,\mat Y_{22}
                +\sin(2v+2\varpi-2\Omega)\,\mat Y_{2-2}\Big)\right)\,.
\end{equation}
Using the Hansen coefficients $X_k^{n.m}(e)$ defined by
\begin{equation}
\left(\frac{r}{a}\right)^n \e^{\ii m v} = \sum_{k=-\infty}^\infty X_k^{n,m}(e) \e^{\ii k {\cal M}} ,
\end{equation}
we finally get
\begin{equation}
\begin{split}
\mat S' = & \sum_{k=-\infty}^\infty \Bigg(
-\frac{1}{3}X_k^{-3,0}(e) \cos(k{\cal M})\,\mat Y_{20}
\\ &
+\frac{1}{6}X_k^{-3,2}(e)\Big(
  \cos(k{\cal M}+2\varpi-2\Omega)\,\mat Y_{22}
+ \sin(k{\cal M}+2\varpi-2\Omega)\,\mat Y_{2-2}
\Big)
\Bigg)\ .
\end{split}
\label{eq.matS}
\end{equation}
The expressions (\ref{eq.matB'}) and (\ref{eq.matS}) allow to compute the averaged Hamiltonian
$\bar H'$. To shorten the notation, we introduce the matrices $\mat Y'_{2m}$ standing for
the rotated spherical harmonic matrices $\mat R' \mat Y_{2m} \mat
R^{\prime\mathrm{T}}$. The resulting Hamiltonian is split into $\bar H' = \bar H'_0 + \bar H'_1$
where $\bar H'_0$ is autonomous whereas $\bar H'_1$ still depends on time. We get
\begin{subequations}
\begin{equation}
\begin{split}
\bar H'_0 = & 
 \frac{1}{2\sca I_{\rm m}}\left(\left(1+\frac{1}{3}\alpha_{\rm m}\right)|\mat \pi'_{\rm m}|^2 -
\alpha_{\rm m}\langle \mat K', \mat \pi'_{\rm m}\rangle^2\right) 
+ \frac{1}{2\sca I_{\rm c}}\left(\left(1+\frac{1}{3}\alpha_{\rm c}\right)|\mat \pi'_{\rm c}|^2 -
\alpha_{\rm c}\langle \mat K', \mat \pi'_{\rm c}\rangle^2\right) 
\\ &
- \frac{2}{3}\frac{\mathcal Gm_0}{a^3} \sca I\, \Bigg(
 \alpha X_0^{-3,0}(e)\langle\mat Y'_{20},\mat Y_{20}\rangle
+\frac{1}{16}\beta X_{2p}^{-3,2}(e)\left(\langle\mat Y'_{22},\mat Y_{22}\rangle
                                             +\langle\mat Y'_{2-2},\mat Y_{2-2}\rangle\right)
\Bigg)
\\ &
-g\langle\mat k_L,\,\mat \pi'_{\rm m} +\mat \pi'_{\rm c}\rangle
- \omega_p \langle \mat K', \mat \pi'_{\rm m} \rangle
\;,
\end{split}
\label{eq.Hbar0}
\end{equation} 
and
\begin{equation}
\begin{split}
\bar H'_1 = & \frac{2}{3}\frac{{\cal G}m_0}{a^3}\sca I\,\Bigg(
 \frac{1}{2}\cos(2\varpi-2\Omega)
\left(
  \alpha X_0^{-3,2}(e) \langle\mat Y'_{20},\mat Y_{22}\rangle
+\frac{1}{2}\beta X_{2p}^{-3,0}(e) \langle\mat Y'_{22}, \mat Y_{20}\rangle
\right)
\\ &
+\frac{1}{2}\sin(2\varpi-2\Omega)
\left(
  \alpha X_0^{-3,2}(e) \langle\mat Y'_{20},\mat Y_{2-2}\rangle
+\frac{1}{2}\beta X_{2p}^{-3,0}(e) \langle\mat Y'_{2-2}, \mat Y_{20}\rangle
\right)
\\ &
-\frac{1}{16}\cos(4\varpi-4\Omega)\,
\beta X_{-2p}^{-3,2}(e)\left(
  \langle\mat Y'_{22},\mat Y_{22}\rangle - \langle\mat Y'_{2-2},\mat Y_{2-2}\rangle
\right)
\\ &
-\frac{1}{16}\sin(4\varpi-4\Omega)\,
\beta X_{-2p}^{-3,2}(e)\left(
  \langle\mat Y'_{22},\mat Y_{2-2}\rangle + \langle\mat Y'_{2-2},\mat Y_{22}\rangle
\right)
\Bigg) \,.
\end{split}
\end{equation}
\end{subequations}
Notice that all terms appearing in $\bar H'_1$ have been neglected in \citep{Peale69} because for
the spin-orbit resonances considered in {\em ibid} ($p=1$ and $p=\frac{3}{2}$), $\bar H'_1$ is
least of order $e^2\sin^2\theta$ where $\theta$ is the obliquity of the body and both quantities
were assumed small. In this work, we neglect these terms as well although we consider arbitrary
large obliquities. The following is thus only valid for small eccentricities or in cases where
$\bar H'_1$ can be averaged over $\varpi-\Omega$.

\section{Cassini states: location}
\label{sec.Cassini}
\label{sec.coplanar}
\begin{figure}
\begin{center}
\includegraphics[width=0.5\linewidth]{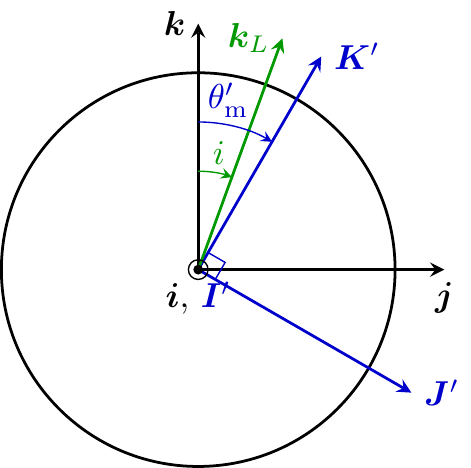}
\caption{\label{fig.conv}Sign convention. $\avec k_L$ is the normal of the invariant plane, $(\avec
I', \avec J', \avec K')$ is a frame attached to the mantle and $(\avec i, \avec j, \avec k)$ is the
orbit frame. When $\avec k$, $\avec k_L$ and $\avec K'$ are coplanar (Cassini's third law), the
orbital inclination $i$ and the obliquity $\theta'_\mathrm{m}$ are taken positive in the clockwise
direction as illustrated in this figure. The same convention applies for the tilt angle of the core
spin-axis $\theta'_\mathrm{c}$.}
\end{center}
\end{figure}

Retaining only $\bar H'_0$ from the Hamiltonian $\bar H'$, the system is said to be in a Cassini
state if and only if the system is at equilibrium in the primed variables, i.e., when all
the following conditions are met simultaneously:
\begin{equation}
\frac{\partial \bar H'_0}{\partial \mat \pi'_{\rm m}} = \mat 0 , \quad
[\frac{\partial \bar H'_0}{\partial \mat\pi'_{\rm c}}, \mat \pi'_{\rm c}] = \mat 0 , \quad
\op J'(\bar H'_0) = \mat 0 \;,
\end{equation}
where (see Appendix~\ref{sec.spin} for the computation of $\op J'(\bar H'_0)$)
\begin{subequations}
\begin{eqnarray}
&&
\label{eq.dHdpis}
\frac{\partial \bar H'_0}{\partial \mat \pi'_{\rm m}} 
  = \frac{1}{\sca I_{\rm m}}\left(\left(1+\frac{1}{3}\alpha_{\rm m}\right)\mat \pi'_{\rm m} -
\alpha_{\rm m}\langle\mat K', \mat \pi'_{\rm m}\rangle\mat K'\right) - g \vec k_L - \omega_p \vec K'
, \\
&&
\label{eq.dHdpio}
[\frac{\partial \bar H'_0}{\partial \mat \pi'_{\rm c}},\mat \pi'_{\rm c} ]
  = -\frac{\alpha_{\rm c}}{\sca I_{\rm c}}\langle\mat K',\mat \pi'_{\rm c}\rangle[\mat K',\mat \pi'_{\rm
c}]
- g[\mat k_L,\mat \pi'_{\rm c}] , \\
&&
\label{eq.JsH}
\begin{split}
\op J'(\bar H'_0) 
  =&
     -\frac{\alpha_{\rm m}}{\sca I_{\rm m}}\langle \mat K',\mat \pi'_{\rm m}\rangle [\mat K', \mat
\pi'_{\rm m}]
     -\frac{\alpha_{\rm c}}{\sca I_{\rm c}}\langle\mat K',\mat \pi'_{\rm c}\rangle[\mat K',\mat \pi'_{\rm
c}]
  -\omega_p[\mat K', \mat \pi'_{\rm m}]
\\ &
    +\frac{2}{3}\frac{\mathcal Gm_0}{a^3}\sca I\,\left(
   \alpha X_0^{-3,0}(e)[\mat Y'_{20},\mat Y_{20}] +\frac{1}{16}
   \beta X_{2p}^{-3,2}(e)\left(
  [\mat Y'_{22},\mat Y_{22}] + [\mat Y'_{2-2},\mat Y_{2-2}]\right)
\right)\;,\qquad
\end{split}\;.
\end{eqnarray}%
\label{eq.dH}%
\end{subequations}%
From Eqs.~(\ref{eq.dHdpis},\ref{eq.dHdpio}), we notice that for the system to be at equilibrium,
$(\avec k_L, \avec K', \avec \pi'_{\rm m}, \avec \pi'_{\rm c})$ have to be coplanar. To verify
Cassini's third law, these vectors must also be coplanar with $\avec k$ (the normal of the orbit
plane). Numerically, we observe that all equilibria do satisfy this condition, therefore we here
limit our quest of the fixed points among those satisfying Cassini's third law.  We further assume
that $\avec i$ and $\avec I'$ are the same vector\footnote{Eventually, we allow the coefficient
$\beta$ to be negative which is equivalent to a rotation of $\pi/2$ around the $K$-axis putting
$\avec J'$ along $\avec i$. This is necessary to stabilise the libration in longitude when
$X_{2p}^{-3,2}(e)$ is negative as in the case of a $1\!:\!2$ spin-orbit resonance, for instance.}.
Within this restriction, the orientations of the mantle and of the core are determined by only two
angles $\theta'_{\rm m}$ and $\theta'_{\rm c}$ representing their respective obliquities.
Furthermore, let $\hmat \pi'_{\rm c}$ be the unit matrix along $\mat \pi'_{\rm c}$.  As said before,
the norm of $\mat\pi'_{\rm c}$ is an integral of the motion and can be set to an arbitrary value.
Hereafter, we assume that $|\mat \pi'_{\rm c}| = C_{\rm c}\omega_p$, whence $\mat \pi'_{\rm c} =
C_{\rm c}\omega_p \hmat \pi'_{\rm c}$. This choice implies that the core is approximately rotating
at the same angular speed as the mantle. Following the common sign convention recalled in
Fig.~\ref{fig.conv}, we have $\hvec \pi'_{\rm c} = \op R_1(-\theta'_{\rm c})\vec k$ and $\mat
R'=\mat R_1(-\theta'_{\rm m})$ leading to $\mat K' = \op R_1(-\theta'_{\rm m})\mat k$.

The conditions for the system to be in a Cassini state can be written in terms of $\theta'_{\rm
m}$ and $\theta'_{\rm c}$ only. In that scope, we notice from Eq.~(\ref{eq.dHdpis}) that
\begin{equation}
\langle \mat K', \mat \pi'_{\rm m} \rangle = C_{\rm m} \big(g\langle\mat K', \mat k_L\rangle + \omega_p \big)\,,\qquad
[\mat K', \mat \pi'_{\rm m}] = \sca I_{\rm m}\left(1-\frac{1}{3}\alpha_{\rm m}\right)g[\mat K', \mat
k_L]\,,
\label{eq.solpim}
\end{equation}
where we used $C_{\rm m} = \sca I_{\rm m}(1+\frac{2}{3}\alpha_{\rm m})$ (see Appendix~\ref{sec.flattening}).
Equalities (\ref{eq.solpim}) allow to get rid of $\mat \pi'_{\rm m}$ in Eqs.~(\ref{eq.dHdpio}, 
\ref{eq.JsH}).
At first order in $\alpha_{\rm c}$, the constraint $[\partial \bar H'_0/\partial \mat \pi'_{\rm c},
\mat \pi'_{\rm c}] = \mat 0$ provides
\begin{subequations}
\begin{equation}
\label{eq.dHdpi0}
\alpha_{\rm c}
\omega_p
  \cos(\theta'_{\rm m}-\theta'_{\rm c})\sin(\theta'_{\rm m}-\theta'_{\rm c})
 + g\sin(i-\theta'_{\rm c}) = 0 \;.
\end{equation}
In the calculation of $\op J'(\bar H'_0)$, the spherical harmonic matrices are rotated according to
Eqs.~(\ref{eq.R1Y2m}) and the commutators $[\mat Y_{2m}, \mat Y_{2k}]$ are taken from the
multiplication table~\ref{tab.mul}. A direct calculation shows that $\op J'(\bar H'_0)$ is
proportional to the matrix $\mat i$. Equating the coefficient to zero, we get at first order in
flattening coefficients:
\begin{equation}
\label{eq.JsH0}
\begin{split}
&
-\alpha_{\rm c} C_{\rm c} \omega_p^2 \cos(\theta'_{\rm m}-\theta'_{\rm c})\sin(\theta'_{\rm
m}-\theta'_{\rm c}) 
\\ &
-\frac{3}{2}\frac{\mathcal Gm_0}{a^3}C\left(
\alpha X_0^{-3,0}(e)\cos\theta'_{\rm m}\sin\theta'_{\rm m}
+\frac{1}{4}\beta X_{2p}^{-3,2}(e)(1+\cos\theta'_{\rm m})\sin\theta'_{\rm m}
\right)
\\ &
-C_{\rm m} g\omega_p\sin(\theta'_{\rm m}-i) = 0 .
\end{split}
\end{equation}%
\label{eq.equilibrium}%
\end{subequations}%
At this stage, it is time to mark a pause to interpret the different terms of
Eqs.~(\ref{eq.equilibrium}) which must be fulfilled by $\theta'_{\rm m}$ and $\theta'_{\rm c}$ for
the system to be at equilibrium. First of all, the last two lines of Eq.~(\ref{eq.JsH0}) are exactly
those obtained by \citet[eq.~18]{Peale69} for a rigid body\footnote{There is a typo in Eq.~(18) of
\citet{Peale69}. A factor 2 is missing before $S(B-A)\sin\theta_1$. From Eq.~(17) of {\em ibid}, we
indeed get $[R(C-\frac{1}{2}A-\frac{1}{2}B)+S(B-A)]\sin2\theta_1+2S(B-A)\sin\theta_1 =
\sin(i-\theta_1)$.  For a direct comparison with the formula of the present paper, let us remind
that $C-(A+B)/2 =\alpha C$, $B-A =\beta C$, and that in the notation of \citet{Peale69},
$R=3n^2\alpha X_0^{-3,0}/(4C\omega_p g)$ and $S=3n^2\beta X_{2p}^{-3,2}/(16C\omega_p g)$.}. Thus in
absence of coupling between the layers, i.e., when the core is spherical ($\alpha_{\rm c}=0$), the
mantle behaves like a rigid planet or satellite as a whole but as if the precession frequency was
reduced by a factor $C_{\rm m}/C$. The remaining term in the first line of Eq.~(\ref{eq.JsH0})
represents the core-mantle coupling. Physically, this is a pressure torque coming from the rotation
of the liquid core. Within the formalism of the present work, this term comes from the kinetic
energy of rotation of the core. Finally, Eq.~(\ref{eq.dHdpi0}) links the pressure torque exerted on
the rigid mantle to the inertial torque induced by the rotating frame. According to this latter
equation, if $\alpha_{\rm c}\rightarrow 0$, then $\sin(i-\theta'_{\rm c}) \rightarrow 0$. Therefore,
in this case the angular momentum of the core tend to be aligned with the normal $\avec k_L$ of the
Laplace plane and is either prograde ($\theta'_{\rm c} = i$) or retrograde ($\theta'_{\rm c} =
i+\pi$).

To simplify a little Eqs.~(\ref{eq.equilibrium}), we apply Kepler's third law $n^2a^3 = {\cal
G}(m_0+M)$ and make the approximation $\omega_p\approx pn$. After some rearrangements, we get
\begin{subequations}
\label{eq.location}
\begin{equation}
\label{eq.dHdpio1}
p \alpha_{\rm c} \cos(\theta'_{\rm m}-\theta'_{\rm c})\sin(\theta'_{\rm m}-\theta'_{\rm c}) + \frac{g}{n} \sin(i-\theta'_{\rm c}) = 0 \;,
\end{equation}
and
\begin{equation}
\label{eq.JsH1}
\begin{split}
&\frac{3}{2p}\frac{m_0}{m_0+M}\frac{n}{g}\left(\alpha X_0^{-3,0}(e)\cos\theta'_{\rm m}\sin\theta'_{\rm
m}
+\frac{1}{4}\beta X_{2p}^{-3,2}(e)(1+\cos\theta'_{\rm m})\sin\theta'_{\rm m}\right)
\\ &
+ \frac{C_{\rm m}}{C}\sin(\theta'_{\rm m}-i) + \left(1-\frac{C_{\rm m}}{C}\right)\sin(\theta'_{\rm c}-i) = 0\;.
\end{split}
\end{equation}%
\end{subequations}%
This set of equations (\ref{eq.location}) is in agreement with Eqs.~(19-21) of \citep{Stys18} when
the moments of inertia of the rigid core in {\em ibid} are set to zero.
In conclusion, the position of the Cassini states are function of eight parameters: the dynamical
flattening coefficients of the whole body $\alpha$, $\beta$, the polar flattening coefficient of the
core $\alpha_{\rm c}$, the polar moment of inertia ratio $C_{\rm m}/C$, the eccentricity $e$, the inclination
$i$, the nodal precession frequency in units of mean motion $g/n$, and the resonance number $p$. For
cases of interest in the solar system, $p$ is either equal to 1 (synchronous rotation as for the
Moon and for most regular satellites of the giant planets) or to $3/2$ ($3\!:\!2$ spin-orbit resonance
like Mercury). The associated Hansen coefficients are
\begin{subequations}
\begin{eqnarray}
&&
X_0^{-3,0}(e) = (1-e^2)^{-3/2}\,,
\\ &&
X_2^{-3,2}(e) = 1-\frac{5}{2}e^2+\frac{13}{16}e^4 + O(e^6)\,, \qquad (p=1)\,,
\\ &&
X_3^{-3,2}(e) = \frac{7}{2}e-\frac{123}{16}e^3+\frac{489}{128}e^5+O(e^7)\,, \qquad (p=3/2)\,.
\end{eqnarray}%
\end{subequations}%
Notice that in situations where the rotation speed is not in resonance with the orbital mean motion
($p$ is not a half integer), $X_{2p}^{-3,2}(e)$ has to be set to zero in Eq.~(\ref{eq.JsH1}).

\section{Cassini states: stability}
\label{sec.stability}
To assert the stability of the Cassini states, we shall linearise the equations of motion in their
vicinity. To do so, we parametrise the state of the system $(\mat \pi'_{\rm m}, \mat
\pi'_{\rm c}, \mat R')$ by a set of 8 coordinates $y' = (\pi'_{{\rm m},x}, \pi'_{{\rm m},y}, \pi'_{{\rm m},z}, \phi'_{\rm c}, \theta'_{\rm c},
\phi'_{\rm m}, \theta'_{\rm m}, \psi'_{\rm m})\in\mathbb{R}^8$ where $(\pi'_{{\rm m},x}, \pi'_{{\rm m},y}, \pi'_{{\rm m},z})$ are the Cartesian coordinates of
the mantle angular momentum, where $(\phi'_{\rm m}, \theta'_{\rm m}, \psi'_{\rm m})$ are the Euler (3,-1,3)
angles defining the rotation matrix $\mat R'$ (i.e., $\mat R' = \mat R_3(\phi'_{\rm m}) \mat
R_1(-\theta'_{\rm m}) \mat R_3(\psi'_{\rm m})$) and where $(\phi'_{\rm c}, \theta'_{\rm c})$ are
defined in such a way that $\hmat \pi'_{\rm c} = \op R_3(\phi'_{\rm c}) \circ \op R_1(-\theta'_{\rm c})\,[\mat k]$.  Within this choice of variables, Cassini states are located in
the subspace $\phi'_{\rm c} = \phi'_{\rm m} = \psi'_{\rm m} = 0$ and ($\theta'_{\rm m},\theta'_{\rm
c}$) have the same meaning as in the previous section, i.e., they correspond to the obliquities of
the two layers with the same sign convention as in \citet{Peale69}. 

Let $y'_e$ be a fixed point of the averaged problem (i.e., the coordinates of a Cassini state), then
$y'_e$ is solution of $\nabla_{y'}\bar H'_0(y'_e)= 0$. The equations of motion are $\dot y' =
-B(y') \nabla_{y'}\bar H'_0(y')$ where $B(y')$ is a Poisson matrix (provided below),
therefore the linearised equations of motion in the vicinity of $y'_e$ are of the form
\begin{equation}
\dot y' = A(y - y'_e)\;,\qquad
A = -B(y'_e) \nabla^2_{y'}\bar H'_0(y'_e)\;.
\label{eq.linear}
\end{equation}

\subsection{Hessian of the Hamiltonian}
To get the Hessian $\nabla^2_{y'}\bar H'_0$ of the Hamiltonian $\bar H'_0$ (\ref{eq.Hbar0}), we
compute its second variation $\delta^2 \bar H'_0$ which is a quadratic form in $\delta y' = (\delta
\pi'_{{\rm m},x}, \delta \pi'_{{\rm m},y}, \delta \pi'_{{\rm m},z}, \delta \phi'_{\rm c}, \delta
\theta'_{\rm c}, \delta \phi'_{\rm m}, \delta \theta'_{\rm m}, \delta \psi'_{\rm m})$. The Hessian
is then the symmetric matrix of this quadratic form. To calculate $\delta^2\bar H'_0$, we use the
fact it is a function of several quantities $(\mat\pi'_{\rm c}, \mat K', \mat Y'_{2m})$ varying
under rotation only. In that scope, we parametrise any infinitesimal rotation $\delta\mat R$
relative to a given rotation $\mat R$ by an element of the respective Lie algebra $\delta\mat\Theta
\equiv \delta\mat R\, \trans{\mat R}$. To the orientation of the core angular momentum
$\mat\pi'_{\rm c}$, represented by the rotation matrix $\mat R_3(\phi'_{\rm c})\mat
R_1(-\theta'_{\rm c})$, we associate the element
\begin{subequations}
\label{eq.dTheta}
\begin{equation}
\delta\mat\Theta'_{\rm c} = \delta\phi'_{\rm c}\mat k - \delta\theta'_{\rm c}\mat i'_{\rm c}\,,
\end{equation}
and for the rotation matrix $\mat R' = \mat R_3(\phi'_{\rm m}) \mat
R_1(-\theta'_{\rm m}) \mat R_3(\psi'_{\rm m})$ we define
\begin{equation}
\delta\mat\Theta'_{\rm m} = \delta\phi'_{\rm m}\mat k - \delta\theta'_{\rm m}\mat i'_{\rm m}
+ \delta\psi'_{\rm m}\mat K'\;.
\label{eq.dThetam}
\end{equation}%
In these equations, $\mat i'_\lambda = {\op R}_3(\phi'_\lambda)[\mat i]$,
$\lambda\in\{{\rm m}, {\rm c}\}$, and $\mat K' = \op R_3(\phi'_{\rm m}) \circ \op R_1(-\theta'_{\rm
m})\,[\mat k]$. In particular, when the system is in a Cassini state, $\mat
i'_{\rm m} = \mat i'_{\rm c} = \mat i$ and $\mat K' = \op R_1(-\theta'_{\rm m})[\mat k]$.
In the derivation of $\delta^2\bar H'_0$, the second variation of the rotation matrices are also
required. We have
\begin{eqnarray}
&&
\delta^2\mat\Theta'_{\rm c} = -\delta\phi'_{\rm c}\delta\theta'_{\rm c}\,[\mat k, \mat i'_{\rm c}]
\,,\\ &&
\delta^2\mat\Theta'_{\rm m} = -\delta\phi'_{\rm m}\delta\theta'_{\rm m}\,[\mat k, \mat i'_{\rm m}]
                          +\delta\phi'_{\rm m}\delta\psi'_{\rm m}\,[\mat k, \mat K']
                          -\delta\theta'_{\rm m}\delta\psi'_{\rm } \,[\mat i'_{\rm m}, \mat K']\,.
\end{eqnarray}%
\end{subequations}%
With these quantities being defined, the first variations of $(\mat\pi'_{\rm c}, \mat K', \mat Y'_{2m})$
are given by
\begin{subequations}
\label{eq.dvar}
\begin{eqnarray}
&&
\label{eq.dpic}
\delta \mat \pi'_{\rm c} = [\delta\mat\Theta'_{\rm c}, \mat \pi'_c]
\,,\\ &&
\delta \mat K' = [\delta \mat \Theta'_{\rm m}, \mat K']
\,,\\ &&
\delta \mat Y'_{2m} = [\delta \mat \Theta'_{\rm m}, \mat Y'_{2m}]\,,
\end{eqnarray}
and their second variations by
\begin{eqnarray}
&&
\delta^2\mat\pi'_{\rm c} = [\delta^2\mat\Theta'_{\rm c}, \mat\pi'_{\rm c}]
                         + [\delta\mat\Theta'_{\rm c}, \delta\mat\pi'_{\rm c}]
\,,\\ &&
\delta^2\mat K' = [\delta^2\mat\Theta'_{\rm m}, \mat K']
                + [\delta\mat\Theta'_{\rm m}, \delta\mat K']
\,,\\ &&
\delta^2\mat Y'_{2m} = [\delta^2\mat\Theta'_{\rm m}, \mat Y'_{2m}]
                     + [\delta\mat\Theta'_{\rm m}, \delta\mat Y'_{2m}]\,.
\end{eqnarray}%
\end{subequations}%
As for the angular momentum of the mantle, we have
\begin{equation}
\label{eq.d12pim}
\delta\mat\pi'_{\rm m} = \delta\pi'_{{\rm m},x}\,\mat i 
+ \delta\pi'_{{\rm m},y}\,\mat j + \delta\pi'_{{\rm m},z}\,\mat k\,,\qquad
\delta^2\mat\pi'_{{\rm m}} = \mat 0\,.
\end{equation}
Furthermore, the Hamiltonian (\ref{eq.Hbar0}) depends on the above variables through scalar products and for any
function $f=\langle \mat A, \mat B \rangle$, we have
\begin{subequations}
\label{eq.dbraket}
\begin{equation}
\delta f = \langle \delta\mat A, \mat B \rangle + \langle \mat A , \delta\mat B \rangle\,,
\qquad
\delta^2f = \langle \delta^2\mat A, \mat B \rangle + 2\langle\delta\mat A,\delta\mat B\rangle +
\langle\mat A,\delta^2\mat B\rangle\,,
\end{equation}
and
\begin{equation}
\delta\left(f^2\right) = 2f\delta f\,,\qquad
\delta^2\left(f^2\right) = 2\left(\delta f\right)^2 + 2f\left(\delta^2f\right)\,.
\end{equation}%
\end{subequations}%
The calculation of $\delta^2\bar H'_0$ has been implemented in TRIP, a general computer algebra
system dedicated to celestial mechanics \citep{Gastineau11}.

\subsection{Poisson matrix}
To get the Poisson matrix $B(y')$ in the variables $y'$ reproducing the equations of motion
(\ref{eq.motion}), we shall first determine the expression of the spin operator $\op J'$. In this
section, we rather consider its vector counterpart $\hat J'$ (i.e., whose image is in
$\mathbb{R}^3$). Let $y'_{\rm m} = (\phi'_{\rm m}, \theta'_{\rm m}, \psi'_{\rm m})$ and $\tens
J'(y'_{\rm m})$ the
matrix such that
\begin{equation}
\hat J'(\bar H'_0) = \tens J'(y'_{\rm m}) \frac{\partial \bar H'_0}{\partial y'_{\rm m}}\,.
\label{eq.JH0}
\end{equation}
By construction, $\tens J'(y'_{\rm m})$ also satisfies \citep[see][]{Boue19}
\begin{equation}
\delta \avec \Theta'_{\rm m} = \left(\trans{\tens J'(y'_{\rm m})}\right)^{-1} \delta y'_{\rm m}\,,
\label{eq.Boue19}
\end{equation}
where $\delta \avec \Theta'_{\rm m}$ is the 3D vector representing the skew-symmetric matrix
$\delta\mat\Theta'_{\rm m}$.  Based on Eq.~(\ref{eq.dThetam}), we get
\begin{equation}
\delta\avec\Theta'_{\rm m} = \begin{pmatrix}
-\delta\theta'_{\rm m}\cos\phi'_{\rm m} - \delta\psi'_{\rm m}\sin\theta'_{\rm m}\sin\phi'_{\rm m} \\
-\delta\theta'_{\rm m}\sin\phi'_{\rm m} + \delta\psi'_{\rm m}\sin\theta'_{\rm m}\cos\phi'_{\rm m} \\
 \delta\phi'_{\rm m} + \delta\psi'_{\rm m}\cos\theta'_{\rm m}
\end{pmatrix}\,,
\end{equation}
and by identification of the last two equations, we obtain\footnote{To avoid fractions, we use the
cotangent and cosecant trigonometric functions respectively defined as $\cot\alpha = (\tan\alpha)^{-1}$ and $\csc \alpha = (\sin\alpha)^{-1}$.}
\begin{equation}
\tens J'(y'_{\rm m}) = \begin{bmatrix}
\cot\theta'_{\rm m}\sin\phi'_{\rm m} && -\cos\phi'_{\rm m} && -\csc\theta'_{\rm m}\sin\phi'_{\rm m} \\
-\cot\theta'_{\rm m}\cos\phi'_{\rm m} && -\sin\phi'_{\rm m} && \csc\theta'_{\rm m}\cos\phi'_{\rm m} \\
1 && 0 && 0
\end{bmatrix} \ .
\label{eq.matJ}
\end{equation}

The first three lines of the Poisson matrix provide the evolution rate of $\mat \pi'_{\rm m}$. From
Eq.~(\ref{eq.pimdot}), we have $\!\!\dot{\ \ \avec \pi^{\,\prime}_{\rm m}} = -\avec\pi'_{\rm m}\times({\partial H'_0}/{\partial \avec\pi'_{\rm m}}) - \hat J' (\bar H'_0)$. As a result, the first three lines of the Poisson matrix are
\begingroup
\begin{equation}
\setlength\arraycolsep{5pt}
\begin{bmatrix}
0 & -\pi'_{{\rm m},z} & \pi'_{{\rm m},y} & 0 & 0 &
  \cot\theta'_{\rm m}\sin\phi'_{\rm m} & -\cos\phi'_{\rm m} & -\csc\theta'_{\rm m}\sin\phi'_{\rm m} \\[0.2em]
\pi'_{{\rm m},z} & 0 & -\pi'_{{\rm m},x} & 0 & 0 &
 -\cot\theta'_{\rm m}\cos\phi'_{\rm m} & -\sin\phi'_{\rm m} & \csc\theta'_{\rm m}\cos\phi'_{\rm m}\\[0.2em]
-\pi'_{{\rm m},y} & \pi'_{{\rm m},x} & 0 & 0 & 0 &
  1 & 0 & 0
\end{bmatrix}\,.
\end{equation}
\endgroup

As for the last three lines of the Poisson matrix, we write the time derivative of $y'_{\rm m}$ in
terms of $\avec\omega'_{\rm m}$ deduced from Eq.~(\ref{eq.Boue19}), namely,
\begin{equation}
\dot y'_{\rm m} = \trans{\tens J'(y'_{\rm m})}\,\avec\omega'_{\rm m}\,,
\label{eq.Boue19b}
\end{equation}
with $\avec\omega_{\rm m} = \partial H'_0/\partial \avec\pi'_{\rm m}$.
Therefore, the last three lines of the Poisson matrix are
\begingroup
\setlength\arraycolsep{5pt}
\begin{equation}
\begin{bmatrix}
-\cot\theta'_{\rm m}\sin\phi'_{\rm m} & \cot\theta'_{\rm m}\cos\phi'_{\rm m} & -1 &
0 & 0 & 0 & 0 & 0 \\
\cos\phi'_{\rm m} & \sin\phi'_{\rm m} & 0 &
0 & 0 & 0 & 0 & 0 \\
\csc\theta'_{\rm m}\sin\phi'_{\rm m} & -\csc\theta'_{\rm m}\cos\phi'_{\rm m} & 0 &
0 & 0 & 0 & 0 & 0
\end{bmatrix}\,.
\label{eq.3lines}
\end{equation}
\endgroup

Finally, to get the entire Poisson matrix, we add the part governing the evolution of the
orientation of the core angular momentum $(\phi'_{\rm c}, \theta'_{\rm c})$. In that scope, we apply
the derivation chain rule.  The result is
\begingroup
\setlength\arraycolsep{1pt}
\begin{equation}
B(y') = 
\begin{bmatrix}
0 & -\pi'_{{\rm m},z} & \pi'_{{\rm m},y} & 0 & 0 &
\displaystyle \frac{\sin\phi'_{\rm m}}{\tan\theta'_{\rm m}} & -\cos\phi'_{\rm m} & \displaystyle-\frac{\sin\phi'_{\rm m}}{\sin\theta'_{\rm m}} \\[1.0em]
\pi'_{{\rm m},z} & 0 & -\pi'_{{\rm m},x} & 0 & 0 &
\displaystyle -\frac{\cos\phi'_{\rm m}}{\tan\theta'_{\rm m}} & -\sin\phi'_{\rm m} & \displaystyle \frac{\cos\phi'_{\rm m}}{\sin\theta'_{\rm m}}\\[1.0em]
-\pi'_{{\rm m},y} & \pi'_{{\rm m},x} & 0 & 0 & 0 &
  1 & 0 & 0 \\[0.2em]
0 & 0 & 0 & 0 & \displaystyle \frac{1}{\pi_c\sin \theta'_{\rm c}} & 0 & 0 & 0 \\
0 & 0 & 0 & \displaystyle -\frac{1}{\pi_c\sin \theta'_{\rm c}} & 0 & 0 & 0 & 0  \\
\displaystyle-\frac{\sin\phi'_{\rm m}}{\tan\theta'_{\rm m}} & \displaystyle\frac{\cos\phi'_{\rm m}}{\tan\theta'_{\rm m}} & -1 &
0 & 0 & 0 & 0 & 0 \\[1.0em]
\cos\phi'_{\rm m} & \sin\phi'_{\rm m} & 0 &
0 & 0 & 0 & 0 & 0 \\[0.4em]
\displaystyle\frac{\sin\phi'_{\rm m}}{\sin\theta'_{\rm m}} & \displaystyle-\frac{\cos\phi'_{\rm m}}{\sin\theta'_{\rm m}} & 0 &
0 & 0 & 0 & 0 & 0
\end{bmatrix}\,.
\end{equation}
\endgroup
We verify that the so-constructed Poisson matrix is skew-symmetric as it should be.

\subsection{Stability criterion}
A fixed point $y'_e$ is said to be {\em Lyapunov stable} if the Hessian $\nabla^2_{y'}\bar H'_0(y'_e)$ is
positive definite (all eigenvalues are strictly positive) or negative definite (all eigenvalues are
strictly negative). The kinetic part of the Hamiltonian guarantees positive eigenvalues and Lyapunov
stable equilibria are de facto associated with a positive definite Hessian. Besides the Lyapunov
stability criterion, which is a strong criterion hardly satisfied in our problem (see
Appendix~\ref{sec.2dof}), we also consider {\em spectral stability} which states that a fixed point
$y'_e$ is stable if all the (complex) eigenvalues of the matrix $A$ of the linearised system
(\ref{eq.linear}) are purely imaginary (no real part). Lyapunov stability implies spectral stability
but the converse is not true. To estimate the impact of dissipation (not included here) on
these two types of stability, we refere the reader to the comprehensive review by
\citet{Krechetnikov07}.

\section{Results}
\label{sec.results}
\subsection{Application to Mercury}
\label{sec.Mercury}
In this section we analyse Mercury's possible Cassini states predicted by the present idealised
model according to which the core is a perfect fluid undergoing Poincar\'e's simple motion and the
mantle is totally rigid.  The physical and orbital parameters, summarised in Table~\ref{tab.param}, are
mainly taken from \citep{Baland17}. The exceptions are the polar moment of inertia ratio $C_{\rm
m}/C$ provided by \citep{Smith12} and the polar flattening of the core $\alpha_{\rm c}$ which we
allow to vary between 0 and $\alpha$.

\begin{table}
\begin{center}
\renewcommand{\arraystretch}{1.5}
\caption{\label{tab.param}Mercury's parameters.}
\begin{tabular}{llll} \hline
Parameter & Symbol & Value & Reference \\ \hline \hline
planet polar flattening coefficient & $\alpha$ & $0.14658\times10^{-3}$ & \citep{Baland17} \\
planet equatorial flattening coefficient &  $\beta$ & $0.93666\times10^{-4}$ & \citep{Baland17} \\
core   polar flattening coefficient &  $\alpha_{\rm c}/\alpha$ & from $\,0\,$ to 1 \\
polar moment of inertia ratio & $C_{\rm m}/C$ & $0.452$ & \citep{Smith12} \\
orbital eccentricity & $e$ & $0.20563$ & \citep{Baland17} \\
orbital inclination & $i$ & $8.533^\circ$ & \citep{Baland17} \\
precession frequency in units of mean motion & $g/n$ & $-0.73990\times10^{-6}$ & \citep{Baland17} \\
resonance & $p$ & $3/2$ \\
planet-to-Sun mass ratio & $M/m_0$ & neglected \\
\hline
\end{tabular}
\end{center}
\end{table}

\begin{figure}
\begin{center}
\includegraphics[width=\linewidth]{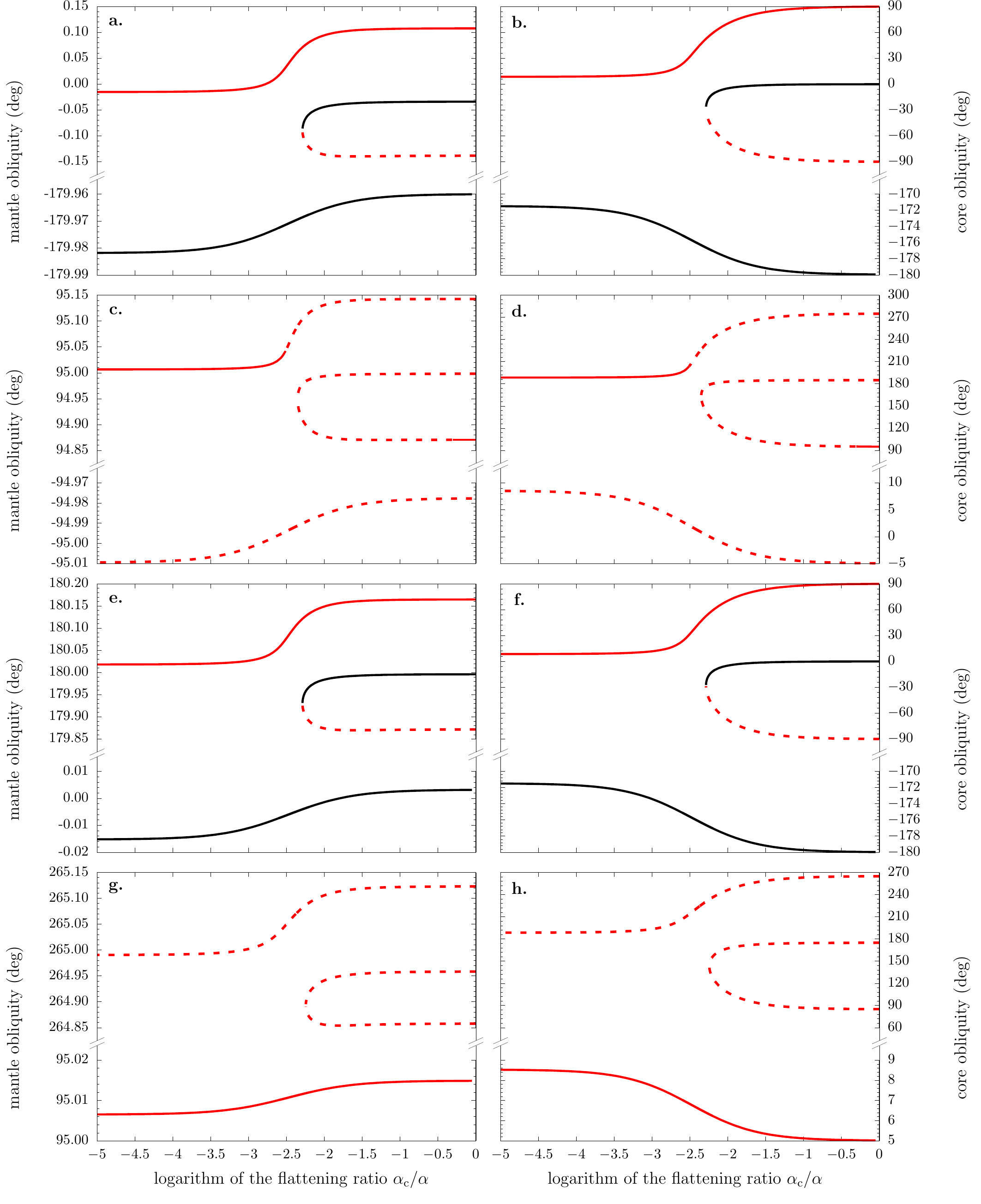}
\caption{\label{fig.Cassini}Mercury's Cassini states as a function of $\log_{10}(\alpha_{\rm
c}/\alpha)$. Each row displays the location of a given set of fixed point families projected onto
$\theta'_{\rm m}$: mantle obliquity (left panel) and $\theta'_{\rm c}$: core obliquity (right
panel). The spectral and Lyapunov stabilities of each Cassini states are determined as described in Sect.~\ref{sec.stability}:
Lyapunov stable equilibria are represented with solid black curves, spectral stable but Lyapunov unstable
ones are plotted with solid red curves, unstable ones are displayed with red dashed curves.
}
\end{center}
\end{figure}

To get the location of Mercury's Cassini states, the system (\ref{eq.location}) is written as a set
of polynomial equations in $\left\{X_{\rm m} = \cos\theta'_{\rm m},\, Y_{\rm m} = \sin\theta'_{\rm
m},\, X_{\rm c} = \cos\theta'_{\rm c},\,Y_{\rm c} = \sin\theta'_{\rm c}\right\}$ to which the
conditions $X_{\rm m}^2 + Y_{\rm m}^2 = 1$ and $X_{\rm c}^2 + Y_{\rm c}^2=1$ are added. The
polynomial system is solved using the command {\tt Isolate} of Maple's package {\tt RootFinding}.
The spectral stability of each solution is determined using TRIP as described in
Section~\ref{sec.stability}. Results are displayed in Figure~\ref{fig.Cassini} as a function of
$\alpha_{\rm c}/\alpha$. Several remarkable points deserve to be commented.

There exists up to 16 different equilibrium solutions. By comparison, a completely
rigid body only possesses at most 4 stationary states \citep{Colombo66, Peale69}.
The solutions presented in Fig.~\ref{fig.Cassini}a and \ref{fig.Cassini}b actually behave like the 4
Cassini states of the classical problem \citep[e.g.,][fig.~3]{Ward04}. Up to a critical value of the
varying parameter (here $\alpha^{\rm crit}_{\rm c}\approx 10^{-2.3}\alpha \approx 0.005\,\alpha$)
there are two elliptical fixed points (solid curves): one prograde close to $0^\circ$ of obliquity and the other
retrograde near $180^\circ$. At $\alpha_{\rm c} = \alpha^{\rm crit}_{\rm c}$ a double fixed point appears.
For $\alpha_{\rm c} > \alpha^{\rm crit}_{\rm c}$, the latter bifurcates into two prograde
equilibria, one being stable (solid black) and the other unstable (dashed red). We also notice that
the bifurcation pattern of the core is significantly more widened than that of the mantle. Whereas
the latter obliquity is bounded between $-0.15^\circ$ and $0.10^\circ$, the core can be tilted up to
$\pm 90^\circ$. Let us emphasise another important point: the Cassini state ranging from $\sim 0^\circ$ to
$\sim0.10^\circ$ is spectrally stable but Lyapunov unstable (solid red curve). This results is quite
puzzling because the Moon actually is in this state (see Section~\ref{sec.low}). Nevertheless, in
Appendix~\ref{sec.2dof} we show that this Lyapunov instability is also predicted in the much more simple
Colombo's top model. We thus conclude that the Lyapunov stability criterion is too restrictive for
our problem and hereafter only consider the spectral stability.

The bifurcation pattern shown in the upper half of Fig.~\ref{fig.Cassini}a is reproduced three times
around $\theta'_{\rm m} = 95^\circ$ (Fig.~\ref{fig.Cassini}c), $180^\circ$
(Fig.~\ref{fig.Cassini}e), and $265^\circ$ (Fig.~\ref{fig.Cassini}g). Moreover, to each of these
equilibrium solutions, there exists an isolated curve of fixed points diametrically opposite, i.e.,
offset by $180^\circ$ (lower half of Figs.~\ref{fig.Cassini}a,c,e,g). Nevertheless there are
discrepancies between the first and the last three rows of Fig.~\ref{fig.Cassini}. In the
latter, the bifurcation pattern of the core is offset from that of the mantle by the order of
$90^\circ$ or $180^\circ$. Therefore, at these new Cassini states, the core is significantly tilted
from the mantle (except for very specific values of the parameter $\alpha_{\rm c}$). Additionally,
the spectral stabilities are not alike in all rows. In the second and fourth rows, we observe
bifurcations with the appearance of two hyperbolic fixed points. Moreover, along a same family of
fixed point, the stability happens to switch from stable to unstable and vice versa
(Figs.~\ref{fig.Cassini}c and \ref{fig.Cassini}d).

When $\alpha_{\rm c} \rightarrow 0$ the core is decoupled from the mantle (see
Eqs.~\ref{eq.location}). Therefore, the latter must behave like a rigid body and cannot have more
than four different Cassini states. This is indeed the case. The four equilibrium obliquities of the
mantle in the limit $\alpha_{\rm c} \rightarrow 0$ are $\theta'_{\rm m} = -0.02^\circ$ and
$180.02^\circ$ (Figs.~\ref{fig.Cassini}a and \ref{fig.Cassini}e), and $\theta'_{\rm m} = \pm
95.01^\circ$ (Figs.~\ref{fig.Cassini}c and \ref{fig.Cassini}g). Moreover, only one of these fixed
points is unstable (the one at $\theta'_{\rm m} = -95.01^\circ$). At each of these four fixed
points, the core obliquity takes two different values, namely $i = 8.533^\circ$ and $i+\pi =
188.533^\circ$, as explained in Sect.~\ref{sec.coplanar}.

In the limit $\alpha_{\rm c} \rightarrow \alpha$, the solution with the lowest obliquity in
Figs.~\ref{fig.Cassini}a and \ref{fig.Cassini}b is at $\theta'_{\rm m} \approx -0.034^\circ$ and
$\theta'_{\rm c} \approx -0.067^\circ$. It is remarkable that for this particular Cassini states,
the core and the mantle are almost aligned. Therefore, the mantle obliquity is close to the value
obtained when assuming the planet to be completely rigid, namely $2.05\,{\rm arcmin} = 0.034^\circ$
\citep[e.g.,][]{Baland17}.

\subsection{Low obliquity Cassini states}
\label{sec.low}
Here we focus on the low obliquity Cassini states of Mercury, the Moon and Io. We choose these
bodies in particular because they all have been described by a model with a liquid core in the
literature. Physical and orbital parameters of Mercury are provided in
Table~\ref{tab.param}, those of the Moon and Io are displayed in Table~\ref{tab.moonio}.

\begin{table}
\begin{center}
\renewcommand{\arraystretch}{1.5}
\caption{\label{tab.moonio}Parameters of the Moon and Io.}
\begin{tabular}{llll} \hline
Parameter & Symbol & Moon & Io \\ \hline \hline
body polar flattening coefficient & $\alpha$ & $0.51690\times10^{-3\;{\rm a}}$  & $4.8982\times10^{-3\;\rm c}$ \\
body equatorial flattening coefficient &  $\beta$ & $0.22772\times10^{-3\;{\rm a}}$ & $5.8770\times10^{-3\;\rm c}$\\
core   polar flattening coefficient &  $\alpha_{\rm c}/\alpha$ & from $\,0\,$ to 1 & from $\,0\,$ to 1 \\
polar moment of inertia ratio & $C_{\rm m}/C$ & $0.9993^{\;\rm a}$ & $0.98346^{\;\rm d}$ \\
orbital eccentricity & $e$ & $0.0549^{\,\rm b}$ & $0.00415^{\;\rm e}$ \\
orbital inclination & $i$ & $5.145^{\circ\;{\rm b}}$ & $0.036^{\circ\;\rm e}$ \\
precession frequency in units of mean motion & $g/n$ & $-0.40188\times10^{-2\;\rm b}$ & $-0.62832\times10^{-3\;\rm e}$\\
resonance & $p$ & $1$ & $1$ \\
body-to-central mass ratio & $M/m_0$ & $0.0123^{\;\rm a}$ & neglected \\
\hline
\end{tabular}
\end{center}
{\footnotesize
$^{\rm a}$ \citep{INPOP17a}; \
$^{\rm b}$ \citep{Yoder95}; \ 
$^{\rm c}$ \citep{Anderson01}; \ 
$^{\rm d}$ \citep[][, Section 4.4, model 1]{Noyelles14}; \ 
$^{\rm e}$ \citep{Lainey06}
}
\end{table}

\begin{figure}
\begin{center}
\includegraphics[width=\linewidth]{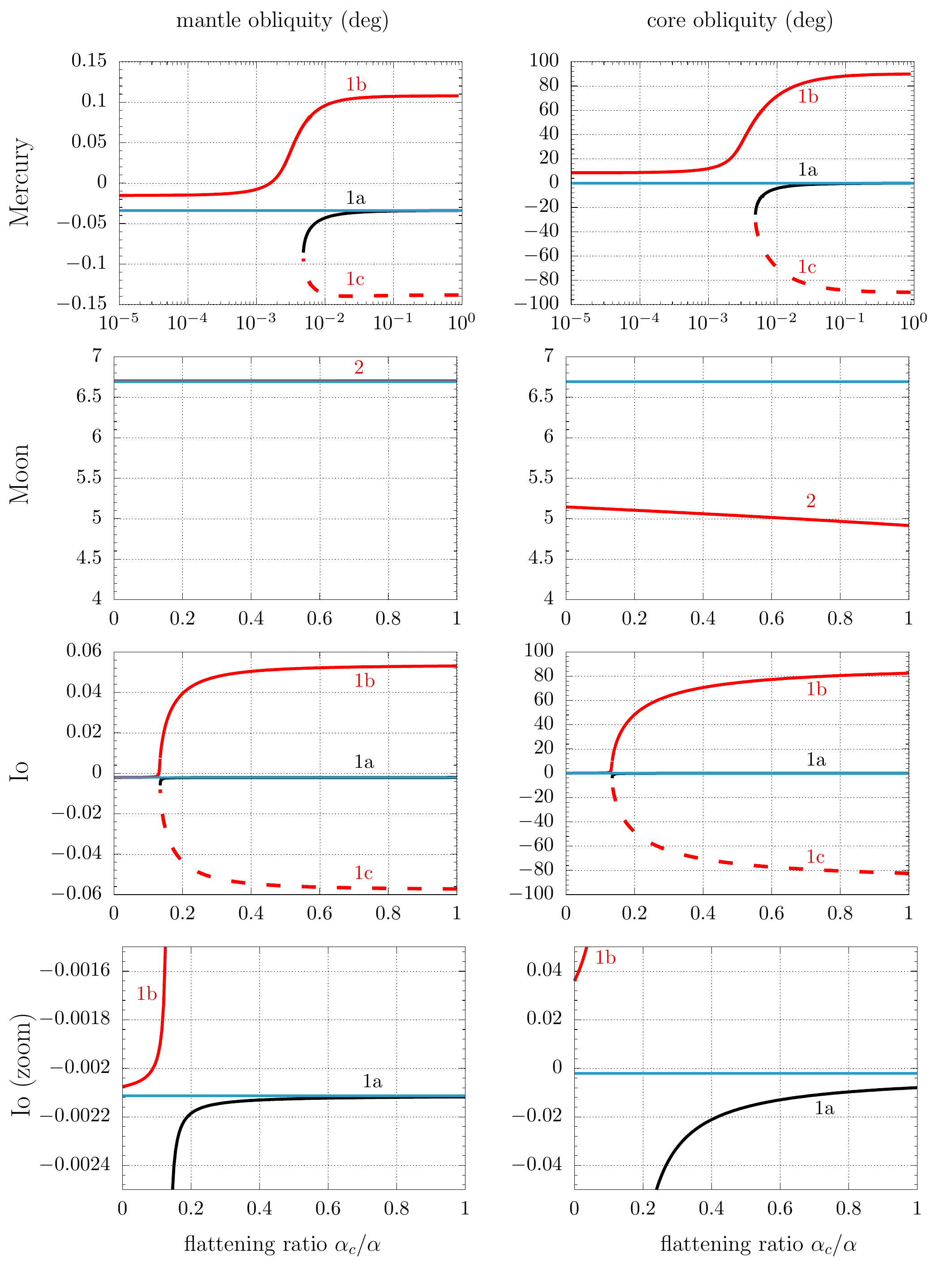}
\caption{\label{fig.lowobliquity}Cassini states labelled `1a', `1b', `1c', `2' as a function of $\alpha_{\rm c}/\alpha$. black
solid curve: spectral and Lyapunov stable points, red solid curves: spectral stable but Lyapunov
unstable points, red dashed curve: spectral unstable points, blue solid curve: solution of the fully
rigid model \citep{Peale69}. Top row: Mercury's Cassini states, second row: Moon's Cassini states,
third row: Io's Cassini states, bottom row: zoom on Io's Cassini states.
}
\end{center}
\end{figure}

The location and stability of the Cassini states are computed as described in
Sect.~\ref{sec.Cassini} and \ref{sec.stability}. These positions are then compared with those
expected from a fully rigid model \citep{Peale69} in Fig.~\ref{fig.lowobliquity}.  According to the
standard rigid body nomenclature of Cassini states \citep{Peale69}, Mercury and Io are lying on
state 1 (the stable branch arising from the bifurcation) while the Moon is settled on state 2 (the
prograde branch extending over the whole parameter space). From Figure~\ref{fig.lowobliquity}, we
observe that for the present values of the Moon's physical and orbital parameters, introducing a
fluid core does not split the branch of Cassini states 2. The derived mantle obliquity is very close
to that predicted by \citeauthor{Peale69}'s formula and the core is only tilted by about
1.5$^\circ$. In the case of Mercury and Io, however, the presence of a liquid core does break state
1 into three different branches labelled `1a', `1b' and `1c' (see Fig.~\ref{fig.lowobliquity}). For
both these bodies, the branch `1a' remain very close to the obliquity inferred from
\citeauthor{Peale69}'s model. When the core flattening coefficient is less than the minimal value
required for the existence of states `1a' and `1c', the branch `1b' is reasonably close the rigid
body expectation. As explained in the previous section, the difference is a function of $C_{\rm
m}/C$. It should be stressed that this offset which implies a departure from the observed obliquity
(in the case of Mercury at least) does not preclude the existence of a spherical core. Indeed, a
slight correction of the physical parameters would allow to recover the observed value. Finally, as
the ratio $\alpha_{\rm c}/\alpha$ goes to 1, the physical validity of states `1a' and `1c' becomes
questionable given the large tilt of the core rotation axis relative to the mantle's figure axis
that reaches 90$^\circ$.

\section{Conclusion}
\label{sec.conclusion}
In this work, we determine the orientations and stabilities of the Cassini states of a rigid body
with a liquid layer described by the Poincar\'e-Hough model. The analysis is performed in a
non-canonical Hamiltonian formalism where variables are represented by matrices. The spin-orbit
gravitational interaction is written in terms of spherical harmonic matrices. The latter have good
mathematical properties allowing to conduct the calculation efficiently.

The problem has four degrees of freedom but the condition for the system to be in a Cassini state is
reduced to a set of only two equations whose unknowns are the mantle obliquity
$\theta_\mathrm{m}$ and the tilt angle of the core spin-axis $\theta_\mathrm{c}$. These equations
are
\begin{equation*}
p \alpha_{\rm c} \cos(\theta_{\rm m}-\theta_{\rm c})\sin(\theta_{\rm m}-\theta_{\rm c}) + \frac{g}{n} \sin(i-\theta_{\rm c}) = 0 \;,
\end{equation*}
and
\begin{equation*}
\begin{split}
&\frac{3}{2p}\frac{m_0}{m_0+M}\frac{n}{g}\left(\alpha X_0^{-3,0}(e)\cos\theta_{\rm m}\sin\theta_{\rm
m}
+\frac{1}{4}\beta X_{2p}^{-3,2}(e)(1+\cos\theta_{\rm m})\sin\theta_{\rm m}\right)
\\ &
+ \frac{C_{\rm m}}{C}\sin(\theta_{\rm m}-i) + \left(1-\frac{C_{\rm m}}{C}\right)\sin(\theta_{\rm c}-i) = 0\;.
\end{split}
\end{equation*}
They depend on eight parameters, namely, the core polar flattening coefficient $\alpha_c$, the overall
polar and equatorial flattening coefficients $\alpha$ and $\beta$, the polar moment of inertia ratio
$C_\mathrm{m}/C$, the body to companion mass ratio $M/m_0$, the orbital inclination $i$ and
eccentricity $e$, the nodal precession frequency and the rotation speed in units of orbital mean
motion $g/n$ and $p$, respectively. If $p$ is not a half integer, the Hansen coefficient
$X_{2p}^{-3,2}(e)$ has to be set to zero. For an evanescent core, i.e., when $C_\mathrm{m} \rightarrow C$,
we retrieve the equation whose roots are the obliquities of the Cassini states of a fully rigid body.

We present the formulae allowing to determine the Lyapunov and spectral stabilities of these Cassini
states. For this problem, the Lyapunov stability criterion is very stringent as it cannot guarantee
the stability of the current state of the Moon. We explain this outcome on a simplified
axisymmetric problem corresponding to Colombo's top.

The model is applied to Mercury, Io and the Moon. For Mercury we highlight 16 different branches of
Cassini states as a function of $\alpha_\mathrm{c}/\alpha$. Given that the problem has three more
degrees of freedom than in the fully rigid body case, the phase space presents more complex
features. In particular, we observe the appearance of two spectrally unstable families from a same
bifurcation point or transitions from stable to unstable configurations along two families
of fixed points. Among the 16 Cassini states, if $\alpha_\mathrm{c}$ is sufficiently close to
$\alpha$, 8 of them can be spectrally stable. In the same limit, one of the Cassini states inferred
by this model for each of the studied body is very closed to that predicted by the fully rigid
model. In most of the other solutions, the core happens to be significantly tilted with respect to the
mantle. These latter configurations might not survive core-mantle friction. This will be the subject
of a future work.

 \section*{Acknowledgements}

 \noindent
 I would like to thank the ASD team for numerous stimulating discussions and Ars\`ene
Pierrot-Valroff for pointing out some errors in a previous version of this paper.

  ~\\
 {\bf{{Conflict of interest}}} The author hereby states that he has no conflict of interest to declare.\\

\appendix

\section{Flattening coefficients}
\label{sec.flattening}
Let $A\leq B \leq C$ be the principal moments of inertia of a given body. The polar and equatorial flattening
coefficients $\alpha$ and $\beta$ are respectively defined as \citep{Vanhoolst02}
\begin{equation}
\alpha = \frac{C-(A+B)/2}{C}\;,\qquad
\beta  = \frac{B-A}{C}\;.
\end{equation}
We denote by $I = (A+B+C)/3$ the mean moment of inertia. The following relations holds
\begin{equation}
A = I\left(1-\frac{1}{3}\alpha-\frac{1}{2}\beta\right)\;,\qquad
B = I\left(1-\frac{1}{3}\alpha+\frac{1}{2}\beta\right)\;,\qquad
C = I\left(1+\frac{2}{3}\alpha\right)\;.
\end{equation}

\section{Spin operator}
\label{sec.spin}
Let $\mat M$ and $\mat N$ be two matrices and $\mat M' = \mat R \mat M \trans{\mat R}$ where
$\mat R$ is a rotation matrix. Under an infinitesimal rotation increment, $\delta \mat R = \delta
\mat \Theta\mat R$, with $\delta\mat \Theta\in\skew(3)$, the matrix $\mat M'$ is transformed
according to
\begin{equation}
\mat M' + \delta\mat M' = (\mat R + \delta \mat R) \mat M \trans{(\mat R + \delta \mat R)}
= \mat M' + [\delta\mat\Theta, \mat M'] + O(|\delta\mat\Theta|^2)\;.
\end{equation}
Let $f=\langle \mat M', \mat N\rangle$. Under the same infinitesimal rotation, the variation of $f$
is given by
\begin{equation}
\begin{split}
\delta f 
& = \langle [\delta\mat\Theta, \mat M'], \mat N\rangle \\
& = \langle \delta\mat\Theta\mat M', \mat N\rangle - \langle \mat M'\delta\mat\Theta, \mat N\rangle \\
& = \langle \delta\mat\Theta, \mat N \mat M^{\prime \mathrm{T}} \rangle - \langle \delta\mat\Theta, \mat
M^{\prime \mathrm{T}}\mat N \rangle \\
& = \langle \delta\mat\Theta, [\mat N , \mat M^{\prime \mathrm{T}}] \rangle \;.
\end{split}
\end{equation}
But by definition, this variation $\delta f$ can also be written
$\delta f = \langle \delta\mat\Theta, \op J(f) \rangle.$
We thus deduce that $\op J(f) = -[\mat M^{\prime \mathrm{T}}, \mat N]$. In particular, 
if $\mat M=\mat S$ is symmetric ($\mat S = \trans{\mat S}$), then
\begin{equation}
\op J(\langle \mat S', \mat N \rangle) = -[\mat S', \mat N]\;,
\end{equation} 
while if $\mat M = \mat A$ is skew-symmetric ($\mat A = -\trans{\mat A}$),
\begin{equation}
\op J(\langle \mat A', \mat N\rangle) = [\mat A', \mat N]\;.
\end{equation}

\section{Stability of Colombo's top}
Colombo's top is an axisymmetric body whose orientation is determined by a single vector representing
the direction of the figure axis. Both the kinetic and the potential energies can be expressed in
terms of this vector. By consequence there is no need to use the matrix formalism described in the
main text. In this Appendix, we take the standard notation up again and write vectors of
$\mathbb{R}^3$ with bold font.

\subsection{One degree of freedom model}
\label{sec.1dof}
Let $\vec s$ be the figure axis and $\alpha$ the precession constant (not the flattening coefficient).
The
orbit plane of normal $\vec k$ is precessing at constant inclination $i$ around the normal
$\vec k_L$ to the Laplace plane. In the frame rotating at the precession frequency $g<0$, where
$\vec k$ and $\vec k_L = \mat R_1(-i)\,\vec k$ are both constant, the Hamiltonian describing the
evolution of Colombo's top is \citep[e.g.,][]{Ward75}
\begin{equation}
H = -\frac{\alpha}{2}(\vec k\cdot\vec s)^2 - g(\vec k_L\cdot\vec s)\,.
\label{eq.H1dof}
\end{equation}
The related equations of motion are \citep{Colombo66}
\begin{equation}
\dvec s = -\hat{J}(H) = \frac{\partial H}{\partial \vec s}\times\vec s
= -\alpha (\vec k \cdot\vec s)(\vec k \times \vec s) - g (\vec k_L \times\vec s)\,.
\label{eq.ds}
\end{equation}
The phase space of this problem is $\mathscr{M}_1 = \{\vec s \in \mathbb{R}^3, |\vec s| = 1\}$. It is of
dimension 2, therefore this problem has a single degree of freedom. The problem can be parametrised
by two angles $(\phi, \theta)$ such that $\vec s = \mat R_3(\phi)\,\mat R_1(-\theta)\,\vec k$.

Cassini states are solution of $\dvec s = \vec 0$. The left-hand side of Eq.~(\ref{eq.ds}) can
vanish only if $\vec s$ is coplanar with $\vec k$ and $\vec k_L$ (Cassini third law) which implies
$\phi=0$. Setting $\phi=0$ in (\ref{eq.ds}), one retrieve the well-known fact that Cassini states'
obliquities $\theta$ are solution of $\frac{\alpha}{g}\cos\theta\sin\theta + \sin(\theta-i) = 0$
\citep[e.g.,][]{Ward04}.

To ascertain the Lyapunov stability of a Cassini state, we evaluate the second derivative of
the Hamiltonian in the vicinity of that given Cassini state. On this purpose, we set
\begin{equation}
\delta \vec s = \delta\vec\theta \times \vec s
\,,\qquad
\delta^2\vec s =
\delta\vec\theta\times(\delta\vec\theta\times\vec s) + \delta^2\vec\theta \times\vec s\,,
\end{equation}
with
\begin{equation}
\delta\vec\theta = \delta\phi\,\vec k - \delta\theta \,\vec i\,,\qquad
\delta^2\vec\theta = -\,\delta\phi\,\delta\theta\,\vec j\,.
\end{equation}
As a result, we get
\begin{equation}
\delta^2H \;=\; -\alpha\left(
\left(\vec k\cdot\delta\vec s\right)^2 + (\vec k\cdot\vec s)(\vec k\cdot\delta^2\vec s)\right)
-g(\vec k_L\cdot\delta^2\vec s) \;=\;
h_{\theta\theta}\,\delta\theta^2
+ h_{\phi\phi}\,\delta\phi^2\,.
\end{equation}
where
\begin{equation}
h_{\theta\theta} = \alpha\cos2\theta + g\cos(\theta-i)\,,\qquad
h_{\phi\phi} = g \sin\theta\sin i\,.
\label{eq.hh}
\end{equation}
The Hamiltonian is locally positive definite if $h_{\theta\theta}$ and $h_{\phi\phi}$ are both
positive or negative. Besides, in this set of coordinates $\vec y = (\theta, \phi)$, the Poisson matrix
reads
\begin{equation}
B(\vec y) = \frac{1}{\sin \theta} \begin{bmatrix}
0 & 1 \\ -1 & 0
\end{bmatrix}\, .
\end{equation}
Therefore, the eigenvalues $\lambda$ of the linearised equations of motion (\ref{eq.linear}) are the
such that $\lambda^2 = - h_{\theta\theta}h_{\phi\phi}/\sin^2\theta$. It follows that the system is
spectrally stable if and only if $h_{\theta\theta}h_{\phi\phi} > 0$, i.e., if and only if the system
is Lyapunov stable. The two criteria are equivalent. By virtue of the expression of $h_{\phi\phi}$
(\ref{eq.hh}), if $-\pi<\theta<0$ (as is the case for Mercury and Io), then stable equilibrium
states correspond to a minimum of $H$ ($h_{\phi\phi}$ is positive because $g$ is negative), but if
$0<\theta<\pi$ (as is the case for the Moon), then the Cassini state is located on a maximum of $H$
($h_{\phi\phi}$ is negative). In the former case, we shall expect that the addition of a (positive
definite) kinetic energy in the Hamiltonian will make the system Lyapunov unstable. This question is
addressed in the following section.

\subsection{Two degrees of freedom model}
\label{sec.2dof}
Hamiltonian (\ref{eq.H1dof}) is only valid in the gyroscopic approximation. Here we add a simple
term accounting for the kinetic energy such that the Lagrangian of the problem reads
\begin{equation}
\hat L = \frac{1}{2}C\hat\omega^2 + \frac{3}{4}(C-A)n^2(\vec k(t)\cdot\hvec s)^2\,,
\end{equation}
where $\hvec \omega$ is the rotation speed, $n$ the mean motion, $A$ the equatorial moment of
inertia, and $C$ the polar moment of inertia. The Lagrangian is defined up to a constant factor. Let
us divide $\hat L$ by $C$ and only then take the Legendre transform to get the Hamiltonian. Moreover, we
choose units of time such that $n=1$. In that case, the moment is $\hvec\pi = \hvec\omega$ and the
Hamiltonian $\hat H$ in the inertial frame reads
\begin{equation}
\hat H = \frac{\hat\pi^2}{2} - \frac{3}{4}\frac{C-A}{C}(\vec k(t)\cdot\hvec s)^2\,,
\end{equation}
with equations of motion \citep[e.g.,][]{Boue06}
\begin{equation}
\frac{\dd}{\dd t}{\hat{\vec\pi}} 
\;=\;
\frac{\partial\hat H}{\partial\hvec\pi} \times \hvec \pi - \hat J(\hat H)
\;=\;
\frac{\partial\hat H}{\partial\hvec\pi} \times \hvec \pi + 
  \frac{\partial\hat H}{\partial\hvec s} \times \hvec s
\,,\qquad
\frac{\dd}{\dd t}\hvec s \;=\; \frac{\partial \hat H}{\partial\hvec \pi} \times \hvec s\,.
\label{eq.eqmH2dof}
\end{equation}
As in the main text, we apply a change of coordinates $(\hvec\pi, \hvec s) \rightarrow (\vec\pi,
\vec s)$ to study the problem in the frame rotating at the precession frequency $g$, i.e., we set
$(\hvec\pi, \hvec s) = \mat R_3(gt)\,(\vec\pi, \vec s)$. To conserve the form of the equations of
motion (\ref{eq.eqmH2dof}), the new Hamiltonian shall read $H(\vec\pi,\vec s) = \hat
H(\hvec\pi,\hvec s) - g (\vec k_L\cdot\vec \pi)$. Let $\gamma = \frac{3}{2}\frac{C-A}{C}$, we get
\begin{equation}
H = \frac{\pi^2}{2} - \frac{1}{2}\gamma(\vec k\cdot\vec s)^2 - g(\vec k_L\cdot\vec\pi)\,.
\end{equation}
This expression is equivalent to Eq.~(1) of \citet{Ward75}.

The phase of the problem is $\mathscr{M}_2 = \{(\vec\pi, \vec
s)\in\mathbb{R}^3\times\mathbb{R}^3,\,|\vec s| = 1 \text{ and }\, (\vec s\cdot\vec \pi) = c\}$ where
$c$ is a constant. This is a manifold of dimension 4, hence the problem has 2 degrees of freedom.
The second condition in the definition of $\mathscr{M}_2$ makes it hard to define a ``natural'' set
of four coordinates to parametrise the phase space. Instead, we use the redundant statevector $\vec y =
(\vec\pi, \vec s)$ where $\vec s$ is parametrised by $(\phi, \theta)$ as in Sect.~\ref{sec.1dof},
i.e., such that $\vec s = \mat R_3(\phi)\mat R_1(-\theta)\vec k$. For $\vec\pi$, we use the
rectangular coordinates $(\pi_x, \pi_y, \pi_z)$. Because the statevector is redundant, we have to
add a Lagrange multiplier $\mu\in\mathbb{R}$ and we introduce the function $F$ defined as
\begin{equation}
F = H + \mu\,(\vec s\cdot\vec \pi)\,.
\end{equation}
The fixed points of the system are given by $\delta F = 0$ with
\begin{equation}
\delta F = (\vec \pi - g \vec k_L + \mu \vec s)\cdot\delta\vec\pi + \left(-\gamma(\vec k\cdot\vec s)\vec
k + \mu \vec\pi\right)\cdot\delta\vec s\,.
\end{equation}
Hence, $\vec \pi$, $\vec s$ and $\mu$ are solution of
\begin{subequations}
\begin{eqnarray}
&&
\label{eq.condpi}
\vec \pi - g \vec k_L + \mu \vec s = \vec 0\,, \\
&&
\label{eq.conds}
-\gamma(\vec k\cdot\vec s)\vec k + \mu\vec\pi = \vec 0\,, \\
&&
\label{eq.condmu}
\vec s\cdot\vec \pi = c
\end{eqnarray}
\end{subequations}
From (\ref{eq.condpi}) and (\ref{eq.condmu}) one gets $\mu = g(\vec k_L\cdot\vec s) - c$.
Substituting this result in Eq.~(\ref{eq.conds}) leads to
\begin{equation}
-\gamma (\vec k\cdot\vec s) (\vec k\times\vec s) + 
\left(g(\vec k_L\cdot\vec s)-c\right) g(\vec k_L\times\vec s) = \vec 0\,.
\label{eq.newCassini}
\end{equation}
Let $\omega_0 = c - g(\vec k_L\cdot\vec s)$. We define the precession constant $\alpha$ as
$\alpha = \gamma / \omega_0$ (this is a misuse of language since by construction $\alpha$ depends on the
orientation $\vec s$). With this definition, the condition (\ref{eq.newCassini}) becomes identical
to (\ref{eq.ds}). We thus retrieve the usual Cassini states.

To analyse the stability, we compute the second variation of $F$, viz.,
\begin{equation}
\delta^2 F = |\delta\vec\pi|^2 -\gamma\left(
(\vec k\cdot\delta\vec s)^2 + (\vec k\cdot\vec s)(\vec k\cdot\delta^2\vec s)\right)
+ \mu \left(\vec\pi\cdot\delta^2\vec s + 2\delta\vec s\cdot\delta\vec\pi\right)\,.
\end{equation}
Substituting the expression of the Lagrange multiplier $\mu$ in this formula, one gets
\begin{equation}
\frac{\delta^2 F}{\omega_0} = \frac{|\delta\vec\pi|^2}{\omega_0}
- \alpha \left(
(\vec k\cdot\delta\vec s)^2 + (\vec k\cdot\vec s)(\vec k\cdot\delta^2\vec s)\right)
- g\,(\vec k_L\cdot\delta^2\vec s) - \omega_0\,(\vec s\cdot\delta^2\vec s) - 2\delta\vec
  s\cdot\delta\vec\pi\,.
\end{equation}
Equivalently, the Hessian of $F$ with respect to $\delta\vec y = (\delta\pi_x, \delta\phi, \delta\pi_y, \delta\pi_z, \delta\theta)\in\mathbb{R}^5$ is
\begingroup
\setlength\arraycolsep{5pt}
\begin{equation}
\nabla^2 F = \omega_0\begin{bmatrix}
1/\omega_0 & \sin\theta & 0 & 0 & 0 \\
\sin\theta & g\sin\theta\sin i + \omega_0\sin^2\theta & 0 & 0 & 0 \\
0 & 0 & 1/\omega_0 & 0 & -\cos\theta \\
0 & 0 & 0 & 1/\omega_0 & \sin\theta \\
0 & 0 & -\cos\theta & \sin\theta & \alpha\cos2\theta + g\cos(\theta-i)+\omega_0
\end{bmatrix}\,.
\label{eq.hessF}
\end{equation}
\endgroup
The seemingly odd order of the components of $\delta\vec y$ has been chosen to highlight the block
matrix structure of $\nabla^2 F$.
The Lyapunov stability of the system is guaranteed if and only if the matrix $\mat Q
\nabla^2F \mat Q$ is definite positive or definite negative where $\mat Q$ is the projection matrix onto
the tangent space, i.e., $\mat Q = \Id - |\vec q|^{-2}\vec q\trans{\vec q}$ where $\vec q$ is the
gradient of the Casimir $C=\vec s\cdot\vec\pi$ of the problem \citep[e.g.,][]{Boue17}. We have
\begin{equation}
\delta C = \vec s \cdot \delta\vec \pi + (\vec s \times\vec \pi)\cdot\delta\vec \theta
\end{equation}
with $\vec s\times\vec \pi = g\,(\vec s \times \vec k_L)$ at equilibrium by virtue of (\ref{eq.condpi}).
From the expressions of $\delta\vec\pi$ and $\delta\vec\theta$, one gets
\begin{equation}
\vec q = (0, 0, \sin\theta, \cos\theta, -g\sin(\theta-i))\,.
\end{equation}
At this stage, an important conclusion can be drawn without performing additional calculation. Let
us decompose the tangent space of the phase space into two linear subspaces $V_1$ and $V_2$ defined
as $V_1 = \{\delta\vec y\in\mathbb{R}^5, \delta\pi_y = \delta\pi_z = \delta\theta = 0\}$ and $V_2 =
\{\delta\vec y\in\mathbb{R}^5, \delta\pi_x = \delta\phi = 0\}$.  The vector $\vec q$ belongs to
$V_2$, therefore the projection matrix $\mat Q$ only acts on $V_2$.  By consequence the submatrix $\mat
F_1$ of $\nabla^2F$ corresponding to the subspace $V_1$ is left unchanged by $\mat Q$. The product
of the eigenvalues of $\mat F_1$ is equal to $\det\mat F_1 = g\omega_0\sin\theta\sin i$.
With $\omega_0>0$ (i.e., $\vec s$ is chosen to point in the same direction as $\vec\pi$), $\det\mat
F_1<0$ when $0<\theta<\pi$. As a result, the system cannot be Lyapunov stable as long as
$0<\theta<\pi$. This is in particular the situation of the Moon. Nevertheless the orientation of
the Moon does not show any sign of instability. We thus conclude that the Lyapunov stability
criterion is too stringent for this problem.

\bibliographystyle{cemda}         
\bibliography{cassini}   

\begin{thebibliography}{50}
\providecommand{\natexlab}[1]{#1}
\providecommand{\url}[1]{{#1}}
\providecommand{\urlprefix}{URL }
\expandafter\ifx\csname urlstyle\endcsname\relax
  \providecommand{\doi}[1]{DOI~\discretionary{}{}{}#1}\else
  \providecommand{\doi}{DOI~\discretionary{}{}{}\begingroup
  \urlstyle{rm}\Url}\fi
\providecommand{\eprint}[2][]{\url{#2}}
\footnotesize

\bibitem[{{Anderson} et~al.(2001){Anderson}, {Jacobson}, {Lau}, {Moore}, and
  {Schubert}}]{Anderson01}
{Anderson} J.~D., {Jacobson} R.~A., {Lau} E.~L., et~al.: {Io's gravity field
  and interior structure}. \jgr {\bfseries 106}(E12), 32,963--32,970 (2001)

\bibitem[{{Baland} et~al.(2011){Baland}, {van Hoolst}, {Yseboodt}, and
  {Karatekin}}]{Baland11}
{Baland} R.-M., {van Hoolst} T., {Yseboodt} M., {Karatekin} {\"O}.: {Titan's
  obliquity as evidence of a subsurface ocean?} \aap {\bfseries 530}:A141
  (2011)

\bibitem[{{Baland} et~al.(2014){Baland}, {Tobie}, {Lef{\`e}vre}, and {Van
  Hoolst}}]{Baland14}
{Baland} R.-M., {Tobie} G., {Lef{\`e}vre} A., {Van Hoolst} T.: {Titan's
  internal structure inferred from its gravity field, shape, and rotation
  state}. \icarus {\bfseries 237}, 29--41 (2014)

\bibitem[{{Baland} et~al.(2017){Baland}, {Yseboodt}, {Rivoldini}, and {Van
  Hoolst}}]{Baland17}
{Baland} R.-M., {Yseboodt} M., {Rivoldini} A., {Van Hoolst} T.: {Obliquity of
  Mercury: Influence of the precession of the pericenter and of tides}. \icarus
  {\bfseries 291}, 136--159 (2017)

\bibitem[{{Beletsky}(2001)}]{Beletsky01}
{Beletsky} V.~V.: {Essays on the Motion of Celestial Bodies}. {Springer Bassel
  AG} (2001)

\bibitem[{{Bills} and {Nimmo}(2011)}]{Bills11}
{Bills} B.~G., {Nimmo} F.: {Rotational dynamics and internal structure of
  Titan}. \icarus {\bfseries 214}, 351--355 (2011)

\bibitem[{{Bou{\'e}}(2017)}]{Boue17b}
{Bou{\'e}} G.: {The two rigid body interaction using angular momentum theory
  formulae}. Celestial Mechanics and Dynamical Astronomy {\bfseries 128}(2-3),
  261--273 (2017)

\bibitem[{{Bou{\'e}} and {Efroimsky}(2019)}]{Boue19}
{Bou{\'e}} G., {Efroimsky} M.: {Tidal Evolution of the Keplerian Elements}.
  Celestial Mechanics and Dynamical Astronomy {\bfseries } (2019)

\bibitem[{{Bou{\'e}} and {Laskar}(2006)}]{Boue06}
{Bou{\'e}} G., {Laskar} J.: {Precession of a planet with a satellite}. \icarus
  {\bfseries 185}(2), 312--330 (2006)

\bibitem[{{Bou{\'e}} and {Laskar}(2009)}]{Boue09}
{Bou{\'e}} G., {Laskar} J.: {Spin axis evolution of two interacting bodies}.
  \icarus {\bfseries 201}(2), 750--767 (2009)

\bibitem[{{Bou{\'e}} et~al.(2009){Bou{\'e}}, {Laskar}, and
  {Kuchynka}}]{Boue09b}
{Bou{\'e}} G., {Laskar} J., {Kuchynka} P.: {Speed Limit on Neptune Migration
  Imposed by Saturn Tilting}. \apjl {\bfseries 702}(1), L19--L22 (2009)

\bibitem[{{Bou{\'e}} et~al.(2016){Bou{\'e}}, {Correia}, and {Laskar}}]{Boue16}
{Bou{\'e}} G., {Correia} A. C.~M., {Laskar} J.: {Complete spin and orbital
  evolution of close-in bodies using a Maxwell viscoelastic rheology}.
  Celestial Mechanics and Dynamical Astronomy {\bfseries 126}(1-3), 31--60
  (2016)

\bibitem[{{Bou{\'e}} et~al.(2017){Bou{\'e}}, {Rambaux}, and {Richard}}]{Boue17}
{Bou{\'e}} G., {Rambaux} N., {Richard} A.: {Rotation of a rigid satellite with
  a fluid component: a new light onto Titan's obliquity}. Celestial Mechanics
  and Dynamical Astronomy {\bfseries } (2017)

\bibitem[{{Bouquillon} et~al.(2003){Bouquillon}, {Kinoshita}, and
  {Souchay}}]{Bouquillon03}
{Bouquillon} S., {Kinoshita} H., {Souchay} J.: {Extension of Cassini's Laws}.
  Celestial Mechanics and Dynamical Astronomy {\bfseries 86}(1), 29--57 (2003)

\bibitem[{{Brasser} and {Lee}(2015)}]{Brasser15}
{Brasser} R., {Lee} M.~H.: {Tilting Saturn without Tilting Jupiter: Constraints
  on Giant Planet Migration}. \aj {\bfseries 150}(5):157 (2015)

\bibitem[{{Cassini}(1693)}]{Cassini93}
{Cassini} G.~D. (1693) {De l'origine et du progr{\`e}s de l'astronomie et de
  son usage dans la g{\'e}ographie et dans la navigation}. In: {Recueil
  d'observations faites en plusieurs voyages par ordre de sa Majest{\'e} pour
  perfectionner l'astronomie et la g{\'e}ographie}, Imprimerie Royale,
  \urlprefix\url{http://dx.doi.org/10.3931/e-rara-7547}

\bibitem[{{Colombo}(1966)}]{Colombo66}
{Colombo} G.: {Cassini's second and third laws}. \aj {\bfseries 71}, 891 (1966)

\bibitem[{{Dufey} et~al.(2009){Dufey}, {Noyelles}, {Rambaux}, and
  {Lemaitre}}]{Dufey09}
{Dufey} J., {Noyelles} B., {Rambaux} N., {Lemaitre} A.: {Latitudinal librations
  of Mercury with a fluid core}. \icarus {\bfseries 203}(1), 1--12 (2009)

\bibitem[{Gastineau and Laskar(2011)}]{Gastineau11}
Gastineau M., Laskar J.: Trip: A computer algebra system dedicated to celestial
  mechanics and perturbation series. ACM Commun Comput Algebra {\bfseries
  44}(3/4), 194--197,
  \urlprefix\url{http://doi.acm.org/10.1145/1940475.1940518} (2011)

\bibitem[{{Hamilton} and {Ward}(2004)}]{Hamilton04}
{Hamilton} D.~P., {Ward} W.~R.: {Tilting Saturn. II. Numerical Model}. \aj
  {\bfseries 128}(5), 2510--2517 (2004)

\bibitem[{{Henrard}(2008)}]{Henrard08}
{Henrard} J.: {The rotation of Io with a liquid core}. Celestial Mechanics and
  Dynamical Astronomy {\bfseries 101}, 1--12 (2008)

\bibitem[{{Hough}(1895)}]{Hough95}
{Hough} S.~S.: {The oscillations of a rotating ellipsoidal shell containing
  fluid}. Philosophical Transactions of the Royal Society of London {\bfseries
  186}, 469--506 (1895)

\bibitem[{{Joachimiak} and {Maciejewski}(2012)}]{Joachimiak12}
{Joachimiak} T., {Maciejewski} A.~J.: {Modeling Precessional Motion of Neutron
  Stars}. In: {Lewandowski} W., {Maron} O., {Kijak} J. (eds) Electromagnetic
  Radiation from Pulsars and Magnetars, Astronomical Society of the Pacific
  Conference Series, vol {\bfseries{}466}, p 183 (2012)

\bibitem[{{Krechetnikov} and {Marsden}(2007)}]{Krechetnikov07}
{Krechetnikov} R., {Marsden} J.~E.: {Dissipation-induced instabilities in
  finite dimensions}. Reviews of Modern Physics {\bfseries 79}(2), 519--553
  (2007)

\bibitem[{{Lainey} et~al.(2006){Lainey}, {Duriez}, and {Vienne}}]{Lainey06}
{Lainey} V., {Duriez} L., {Vienne} A.: {Synthetic representation of the
  Galilean satellites' orbital motions from L1 ephemerides}. \aap {\bfseries
  456}(2), 783--788 (2006)

\bibitem[{{Meyer} and {Wisdom}(2011)}]{Meyer11}
{Meyer} J., {Wisdom} J.: {Precession of the lunar core}. \icarus {\bfseries
  211}(1), 921--924 (2011)

\bibitem[{{Noyelles}(2012)}]{Noyelles12}
{Noyelles} B.: {Behavior of nearby synchronous rotations of a
  Poincar{\'e}-Hough satellite at low eccentricity}. Celestial Mechanics and
  Dynamical Astronomy {\bfseries 112}, 353--383 (2012)

\bibitem[{{Noyelles}(2014)}]{Noyelles14}
{Noyelles} B.: {Contribution {\`a} l'{\'e}tude de la rotation r{\'e}sonnante
  dans le Syst{\`e}me Solaire (in French)}. Habilitation Thesis,
  \urlprefix\url{{https://arxiv.org/abs/1502.01472}} (2014)

\bibitem[{{Noyelles} and {Nimmo}(2014)}]{Noyelles14b}
{Noyelles} B., {Nimmo} F.: {New clues on the interior of Titan from its
  rotation state}. In: IAU Symposium, IAU Symposium, vol {\bfseries{}310}, pp
  17--20 (2014)

\bibitem[{{Noyelles} et~al.(2010){Noyelles}, {Dufey}, and
  {Lemaitre}}]{Noyelles10}
{Noyelles} B., {Dufey} J., {Lemaitre} A.: {Core-mantle interactions for
  Mercury}. \mnras {\bfseries 407}(1), 479--496 (2010)

\bibitem[{{Peale}(1969)}]{Peale69}
{Peale} S.~J.: {Generalized Cassini's Laws}. \aj {\bfseries 74}, 483 (1969)

\bibitem[{{Peale}(1976)}]{Peale76}
{Peale} S.~J.: {Does Mercury have a molten core?} \nat {\bfseries 262}(5571),
  765--766 (1976)

\bibitem[{{Peale} et~al.(2014){Peale}, {Margot}, {Hauck}, and
  {Solomon}}]{Peale14}
{Peale} S.~J., {Margot} J.-L., {Hauck} S.~A., {Solomon} S.~C.: {Effect of
  core-mantle and tidal torques on Mercury{\textquoteright}s spin axis
  orientation}. \icarus {\bfseries 231}, 206--220 (2014)

\bibitem[{{Poincar{\'e}}(1910)}]{Poincare10}
{Poincar{\'e}} H.: {Sur la pr{\'e}cession des corps d{\'e}formables}. Bulletin
  Astronomique {\bfseries 27}, 321--357 (1910)

\bibitem[{{Quillen} et~al.(2018){Quillen}, {Chen}, {Noyelles}, and
  {Loane}}]{Quillen18}
{Quillen} A.~C., {Chen} Y.-Y., {Noyelles} B., {Loane} S.: {Tilting Styx and Nix
  but not Uranus with a Spin-Precession-Mean-motion resonance}. Celestial
  Mechanics and Dynamical Astronomy {\bfseries 130}(2):11 (2018)

\bibitem[{{Ragazzo} and {Ruiz}(2015)}]{Ragazzo15}
{Ragazzo} C., {Ruiz} L.~S.: {Dynamics of an isolated, viscoelastic,
  self-gravitating body}. Celestial Mechanics and Dynamical Astronomy
  {\bfseries 122}, 303--332 (2015)

\bibitem[{{Ragazzo} and {Ruiz}(2017)}]{Ragazzo17}
{Ragazzo} C., {Ruiz} L.~S.: {Viscoelastic tides: models for use in Celestial
  Mechanics}. Celestial Mechanics and Dynamical Astronomy {\bfseries 128},
  19--59 (2017)

\bibitem[{{Smith} et~al.(2012){Smith}, {Zuber}, {Phillips}, {Solomon}, {Hauck},
  {Lemoine}, {Mazarico}, {Neumann}, {Peale}, {Margot}, {Johnson}, {Torrence},
  {Perry}, {Rowlands}, {Goossens}, {Head}, and {Taylor}}]{Smith12}
{Smith} D.~E., {Zuber} M.~T., {Phillips} R.~J., et~al.: {Gravity Field and
  Internal Structure of Mercury from MESSENGER}. Science {\bfseries 336}, 214
  (2012)

\bibitem[{{Stys} and {Dumberry}(2018)}]{Stys18}
{Stys} C., {Dumberry} M.: {The Cassini State of the Moon's Inner Core}. Journal
  of Geophysical Research (Planets) {\bfseries 123}(11), 2868--2892 (2018)

\bibitem[{{Tisserand}(1891)}]{Tisserand91}
{Tisserand} F.: {Trait{\'e} de m{\'e}canique c{\'e}leste. Th{\'e}orie de la
  figure des corps c{\'e}lestes et de leur mouvement de rotation},
  {Gauthier-Villars et fils (Paris)}, chap {XXVIII. Libration de la Lune}.
  \urlprefix\url{https://gallica.bnf.fr/ark:/12148/bpt6k6537806n/f464.image}
  (1891)

\bibitem[{{Touma} and {Wisdom}(2001)}]{Touma01}
{Touma} J., {Wisdom} J.: {Nonlinear Core-Mantle Coupling}. \aj {\bfseries
  122}(2), 1030--1050 (2001)

\bibitem[{{Van Hoolst} and {Dehant}(2002)}]{Vanhoolst02}
{Van Hoolst} T., {Dehant} V.: {Influence of triaxiality and second-order terms
  in flattenings on the rotation of terrestrial planets. I. Formalism and
  rotational normal modes}. Physics of the Earth and Planetary Interiors
  {\bfseries 134}, 17--33 (2002)

\bibitem[{{Van Hoolst} et~al.(2009){Van Hoolst}, {Rambaux}, {Karatekin}, and
  {Baland}}]{Vanhoolst09}
{Van Hoolst} T., {Rambaux} N., {Karatekin} {\"O}., {Baland} R.-M.: {The effect
  of gravitational and pressure torques on Titan's length-of-day variations}.
  \icarus {\bfseries 200}, 256--264 (2009)

\bibitem[{{Viswanathan} et~al.(2017){Viswanathan}, {Fienga}, {Gastineau}, and
  {Laskar}}]{INPOP17a}
{Viswanathan} V., {Fienga} A., {Gastineau} M., {Laskar} J. (2017) {INPOP17a
  planetary ephemerides scientific notes}. Tech. rep.

\bibitem[{{Vokrouhlick{\'y}} and {Nesvorn{\'y}}(2015)}]{Vokroulicky15}
{Vokrouhlick{\'y}} D., {Nesvorn{\'y}} D.: {Tilting Jupiter (a bit) and Saturn
  (a lot) during Planetary Migration}. \apj {\bfseries 806}(1):143 (2015)

\bibitem[{{Ward}(1975)}]{Ward75}
{Ward} W.~R.: {Tidal friction and generalized Cassini's laws in the solar
  system.} \aj {\bfseries 80}, 64--70 (1975)

\bibitem[{{Ward} and {Canup}(2006)}]{Ward06}
{Ward} W.~R., {Canup} R.~M.: {The Obliquity of Jupiter}. \apjl {\bfseries
  640}(1), L91--L94 (2006)

\bibitem[{{Ward} and {Hamilton}(2004)}]{Ward04}
{Ward} W.~R., {Hamilton} D.~P.: {Tilting Saturn. I. Analytic Model}. \aj
  {\bfseries 128}, 2501--2509 (2004)

\bibitem[{{Williams} et~al.(2001){Williams}, {Boggs}, {Yoder}, {Ratcliff}, and
  {Dickey}}]{Williams01}
{Williams} J.~G., {Boggs} D.~H., {Yoder} C.~F., et~al.: {Lunar rotational
  dissipation in solid body and molten core}. \jgr {\bfseries 106}(E11),
  27,933--27,968 (2001)

\bibitem[{{Yoder}(1995)}]{Yoder95}
{Yoder} C.~F.: {Astrometric and Geodetic Properties of Earth and the Solar
  System}. In: {Ahrens} T.~J. (ed) Global Earth Physics: A Handbook of Physical
  Constants, p~1 (1995)

\end{thebibliography}

\end{document}